\documentclass[twocolumn,showpacs,amsmath,amssymb,floatfix,aps,prb,superscriptaddress]{revtex4-1}

\newcommand{\bk}{{\bf k}}

\newcommand{\ba}{{\bf a}}

\newcommand{\br}{{\bf x}}

\newcommand{\beq}{\begin{eqnarray}}
\newcommand{\eeq}{\end{eqnarray}}
\newcommand{\bse}{\begin{subequations}}
\newcommand{\ese}{\end{subequations}}
\newcommand{\beqq}{\begin{eqnarray*}}
\newcommand{\eeqq}{\end{eqnarray*}}

\newcommand{\be}{\begin{equation}}
\newcommand{\ee}{\end{equation}}
\newcommand{\oh}{{\frac{1}{2}}}

\newcommand{\ve}{{\varepsilon}}
\newcommand{\tr}{\tilde r}

\def \nn{\nonumber}

\def \br{{\bf x}}

\def \kv{{\bf k}}
\def \bk{{\bf k}}

\def \bx{{\bf x}}

\def \a{{\alpha}}

\def \g{{\gamma}}

\def \s{{\sigma}}

\def \ve{{\varepsilon}}

\def \ba{\begin{align*}}
\def \ea{\end{align*}}

\newcounter{indice}

\def \mc{\mathcal}

\usepackage{textcomp}
\usepackage[colorlinks=true,citecolor=blue,linkcolor=blue]{hyperref}
\usepackage{graphicx}
\usepackage{tabularx}

\begin{document}

\begin{titlepage}

\title{Ferromagnetic transition in a one-dimensional spin-orbit-coupled metal and its mapping to a critical point in smectic liquid crystals}

\author{Vladyslav Kozii}
\affiliation{Department of Physics, Massachusetts Institute of Technology, Cambridge,
Massachusetts 02139, USA}
\author{Jonathan Ruhman}
\affiliation{Department of Physics, Massachusetts Institute of Technology, Cambridge,
Massachusetts 02139, USA}
\author{Liang Fu}
\affiliation{Department of Physics, Massachusetts Institute of Technology, Cambridge,
Massachusetts 02139, USA}
\author{Leo Radzihovsky}
\affiliation{Department of Physics, Center for Theory of Quantum Matter, and JILA, University of Colorado, Boulder, Colorado 80309, USA}
\affiliation{
Kavli Institute for Theoretical Physics, University of California, Santa Barbara, California 93106, USA}

\date{\today}

\begin{abstract}

We study the quantum phase transition between a paramagnetic and ferromagnetic metal in the presence of Rashba spin-orbit coupling in one dimension. Using bosonization, we analyze the transition by means of renormalization group, controlled by an $\ve$-expansion around the upper critical dimension of two. We show that the presence of Rashba spin-orbit coupling allows for a new nonlinear term in the bosonized action, which generically leads to a fluctuation driven first-order transition.
We further demonstrate that the Euclidean action of this system maps onto a classical smectic-A -- C phase transition in a magnetic field in two dimensions. We show that the smectic transition is second order and is controlled by a new critical point.

\end{abstract}


\maketitle

\draft

\vspace{2mm}

\end{titlepage}

\section{Introduction}

\subsection{Background and motivation}

Quantum phase transitions continue to be one of the central topics in condensed matter physics. The problem is
especially challenging in systems of itinerant electrons. The
first attempts to describe the critical behavior of interacting
itinerant electrons were made by Hertz \cite{Hertz76} and Millis
\cite{Millis93} in the context of the ferromagnetic (FM) and
antiferromagnetic phase transitions. They constructed an effective
Ginzburg-Landau theory by integrating out fermionic degrees of
freedom. However, non-analyticities generated by integrating out
gapless electrons call into question the validity of this uncontrolled
approach. To avoid these dangerous singularities, the gapless fermions and
the soft bosonic order parameter must be treated on equal footing
\cite{BelitzKirkpatrick20011,BelitzKirkpatrick20012,Abanov03,Rech06}. However, the theory still exhibits a divergent
perturbation theory associated with gappless Fermi surface degrees of
freedom, whose control remains an open problem
\cite{BelitzKirkpatrick02,KirkpatrickBelitz03,Chubukov04,Lee09,MetlitskiSachdev,Mross10,Fitzpatrick13,Brandoetal16}.



In contrast, in one dimension, bosonization of the electronic
quasiparticles significantly simplifies the problem, making it
tractable. This approach, combined with a renormalization group (RG)
analysis, was successfully applied by Yang~\cite{Yang04} to analyze the
quantum transition from  paramagnetic (PM) phase to an Ising itinerant ferromagnet in a one-dimensional conductor. The resulting strongly-interacting critical
point is distinct from the Luttinger liquid and Ising critical points,
and in one-loop approximation is characterized by the dynamic critical exponent $z=2$. Later, similar results for a one-dimensional Heisenberg ferromagnet were obtained in Ref.~\onlinecite{SenguptaKim05}.

The possibility of a ferromagnetic ground state seemingly contradicts the Lieb and Mattis theorem \cite{LiebMattis62}, which states that an unmagnetized state always has lower energy for certain classes of systems. This theorem, however, does not take into account spin-orbit coupling, which will play an important role in our study. Furthermore, it was shown that the inclusion of further neighbor hopping terms in the lattice models \cite{DaulNoack98}, as well as a spin-dependent interaction, can also stabilize a ferromagnetic ground state. Finally, numerical results obtained in Ref.~\onlinecite{DaulNoack98} suggest an existence of a ferromagnetic transition in one-dimensional systems. Taken together, these arguments demonstrate that Lieb-Mattis theorem is not applicable in the most general system, studied in this work, implying that the problem of a one-dimensional ferromagnetic transition is well-defined and meaningful.

The model of a ferromagnetic transition studied in Ref.~\onlinecite{Yang04} does not take into account generically present spin-orbit coupling, and relies on the presence of inversion symmetry. In systems that lack inversion symmetry, however, the presence of Rashba spin-orbit coupling leads to interesting physical consequences. It naturally reduces the $SU(2)$ spin rotation symmetry to a $U(1)$ symmetry associated with the total $S_z$ conservation, and, as a result, the ferromagnetic transition becomes of the Ising type. This situation is very common in realistic experimental setups with spin-orbit-coupled wires, where Rashba coupling appears, e.g., due to the internal crystal structure or due to external sources, such as the substrate, gates etc.

Motivated by this observation, in this work we consider the most general case of the ferromagnetic transition in spin-orbit-coupled one-dimensional metals, without assuming any other symmetry except time-reversal. We study it via an RG analysis controlled by an $\ve$-expansion. We show that, in the absence of inversion symmetry, non-linear coupling between spin current and magnetization enhances quantum fluctuations, which {\em generically} drive the itinerant ferromagnetic
transition first-order, akin to a compressible Ising
model~\cite{BergmanHalperin1976}. As a special case, when inversion symmetry is present, we recover the continuous transition found in Ref.~\onlinecite{Yang04}.

We further show that the bosonic Euclidean (imaginary time) $D = 1+1$
dimensional action of an itinerant magnetic wire maps onto a field
theory describing a classical two-dimensional ($D=2$) smectic-A to
smectic-C phase transition in a magnetic field, studied at upper critical dimension, $D=3$, by Grinstein and Pelcovits in their seminal work~\cite{GrinsteinPelcovits82}. They showed that, at $D=3$, the transition is controlled by the Gaussian fixed point, and correlation functions exhibit mean-field like behavior with logarithmical corrections to scaling. We reproduce their results in $D=3$. In addition, we analyze the system below the upper critical dimension, $D<3$, where the Gaussian fixed point becomes unstable. We find a new strongly-interacting stable fixed point which controls the second-order transition.


For bare couplings satisfying a special relation, which
corresponds to a rotationally invariant smectic, the model we consider reduces
to that of anomalous elasticity~\cite{GrinsteinPelcovits81} of a two-dimensional smectic-A liquid
crystal. This problem was solved exactly by Golubovic and
Wang~\cite{GolubovicWang92,GolubovicWang94} through mapping onto the $1+1$
dimensional Kardar-Parisi-Zhang equation~\cite{KardarParisiZhang86}. Thus, our theory reproduces results  known in literature as special limiting cases.

The remainder of the paper is organized as follows. We conclude the
Introduction with a summary of our results. In Sec.~\ref{Sec:microscopicmodel},  we present a
microscopic model for the Ising transition in an itinerant ferromagnet,
including the effect of Rashba spin-orbit coupling. Utilizing bosonization we derive an effective low-energy field theory for this transition. In Sec.~\ref{Sec:FMtransition}, by generalizing the field theory to $D=d+1$
dimensions, we analyze this transition using renormalization
group methods, controlled by an $\ve = 3-D$-expansion. In Sec.~\ref{Sec:smectics}, we
apply these results to a mathematically related classical problem of a
smectic-A to smectic-C transition in a magnetic field and obtain a nontrivial critical
point in two dimensions. We summarize our results and conclude in
Sec.~\ref{Sec:conclusion}.

\subsection{Results}

Before presenting technical details we briefly summarize our findings.
We develop a field-theoretic model for a quantum phase transition to
an itinerant ferromagnet in the presence of Rashba spin-orbit coupling. In contrast to a special case of inversion-symmetric system\cite{Yang04}, we show that generically it
exhibits an additional nonlinearity with a coupling $g_1 > 0$, which is
relevant for $D=d+1<3$, and thus the transition is governed by a qualitatively distinct
behavior (see Sec.~\ref{Sec:FMtransition} and Eq.~(\ref{Eq:g1g2}) for the definition of $g_1$).

Utilizing a one-loop RG method, controlled by an $\ve = 3-D$-expansion,
we show that, akin to a compressible Ising
model~\cite{BergmanHalperin1976}, the inversion-symmetry breaking
nonlinearity $g_1 > 0$, together with strong quantum fluctuations,
generically drive the itinerant ferromagnetic transition first-order. While our analysis relies on an analytical continuation of a 1d bosonized model to high dimensions, we conjecture that this fluctuations-driven first-order transition is a qualitative feature that extends to two and three dimensional ferromagnets without inversion symmetry \cite{RuhmanBerg14,Brandoetal16}.


The imaginary time (Euclidean) action of the model characterizes the
classical Sm-A to Sm-C liquid crystal transition in a magnetic field,
extending seminal work of Grinstein and Pelcovits away from the
marginal dimension of $D=3$ down to $D=2$. Specifically we find that
for $g_1 < 0$, within the one-loop $\ve = 3-D$-expansion ($\ve=1$ for the physical
case), the new critical point that controls the transition is characterized by

\begin{align}
&z = 2 - \frac{3\ve}{37} , \qquad &&\nu = \frac12 + \frac{9\ve}{74} , \nonumber \\
&\gamma = 1 + \frac{\ve}{74} ,\qquad &&\beta = \frac12 - \frac{5\ve}{37},
\end{align}
where $z$, $\nu$, $\gamma$, and $\beta$ are dynamical, correlation length, susceptibility, and order parameter critical exponents, respectively.

Within the global phase diagram of the Euclidean field theory the
continuous transition criticality for $g_1 < 0$ is separated from the
first-order fluctuation-driven transition for $g_1>0$ by the $g_1=0$
inversion-symmetric tricritical point\cite{Yang04}.


\section{Microscopic model for 1d itinerant ferromagnetic transition \label{Sec:microscopicmodel}}

We begin with a generic microscopic model of an
one-dimensional metal with Rashba spin-orbit coupling, characterized by an electronic
Hamiltonian (choosing units such that $\hbar = k_B = 1$)

\be
H = H_{0} + H_{\text{so}} + H_{\text{int}}, \label{Eq:H}
\ee

where

\be
H_{0} = \sum_s\int dx \, \psi_s^\dagger(x) \left[ \ve(-i\partial/\partial x) - \ve_F\right] \psi_s(x)
\ee
is a single-particle band Hamiltonian, characterized by a
dispersion $\ve (-i\partial/\partial x)$, Fermi
level $\ve_F$, and $s=\uparrow,\downarrow$ labels electron
spin projection. $H_{\text{int}}$ accounts for forward- and back-scattering processes.

The second contribution,

\be
H_{\text{so}} = \alpha_R \sum_{s, s'} \int dx \, \psi^\dagger_s(x) \sigma^z_{ss'}(-i\partial/\partial x) \psi_{s'}(x),
\ee
is the Rashba spin-orbit coupling, which is odd under inversion. As mentioned above, we expect this term  to apply to many realistic experimental setups, including semiconducting nanowires with strong spin-orbit coupling \cite{mourik2012signatures,das2012zero,krogstrup2015epitaxy} and noncentrosymmetric quasi one-dimensional materials.

To study the ferromagnetic transition, we now derive the corresponding low-energy Hamiltonian. Focusing
on the vicinity of the Fermi points at momenta $\pm k_F$, we expand
the electron field operators
\be
\psi_s(x) \approx \psi_{sR}(x)  e^{ik_F x} + \psi_{sL}(x) e^{-ik_F x}
\ee
in terms of left ($r=L$) and right ($r=R$) moving fields
$\psi_{sr}(x)$ varying slowly on the scale of Fermi wavelength,
$\lambda \sim 1/k_F$, and satisfying usual anti-commutation relations
$\{\psi_{sr}(x), \psi_{s'r'}(x') \} = \delta_{ss'} \delta_{rr'}
\delta(x-x')$. In terms of the 'slow' fields $\psi_{sR(L)},$ Rashba
spin-orbit coupling reduces to
\be
H_{\text{so}} = \alpha_R k_F \sum_{r,s} \int dx \, r\, s \, n_{sr}(x) = \alpha_R k_F \int dx \, J, \label{Eq:H_so}
\ee
where we defined spin density, $n_{sr}(x) = \psi^\dagger_{s r} (x)
\psi_{s r} (x)$ and spin current density, $J$. Both
$s=\uparrow,\downarrow$ and $r = R,L$ correspond to $+1,-1$,
respectively.

In terms of the slow chiral fields the band Hamiltonian in Eq.~(\ref{Eq:H})
reduces to $H_0 = \sum_n H_{(n)}$, with

\be
\left. H_{(n)} = \frac{\partial^{(n)}_k \ve(k)}{n!} \right\vert_{k=k_F} \sum_{r, s}\int dx \psi_{s r}^\dagger(x) ( -i r \partial_x )^n \psi_{s r}(x).
\ee
Terms with $n\ge 2$ originate from the curvature of the band
dispersion around Fermi points $\pm k_F$
and are often neglected. However, they are crucial here to properly
describe the itinerant FM transition.

Equation~(\ref{Eq:H}) with short-range interaction,
$H_{\text{int}} = U \int dx \, n_{\uparrow}(x) n_{\downarrow}(x)$, and
without spin-orbit coupling, $\alpha_R = 0$, is precisely a starting
point for Hertz-Millis analysis in higher dimensions. It is studied by introducing a
Hubbard-Stratonovich magnetization order parameter to decouple the
interaction and by integrating out electrons to obtain the nonlocal
Landau action for the magnetization. Such an analysis, however, has been shown to be oversimplified, since it does not take into account massless fermionic modes carefully, neglecting important non-analyticities appearing after integrating out electrons, and thus unlikely to describe the true critical behavior.

To avoid the aforementioned difficulties with the Hertz-Millis
approach, here we will utilize the power of one-dimensional Abelian
bosonization to derive an effective bosonic theory that we will then
study by conventional Wilsonian RG\cite{WilsonRG}. Following standard
bosonization procedure~\cite{Giamarchi}, we write electronic operators
in terms of bosonic phase fields:
\be
\psi_{sr} = {F_s\over \sqrt{2\pi a}} e^{i r k_F x}e^{{i\over \sqrt{2}}\left( \phi_\rho+s \,\phi_\sigma-r\,\theta_\rho- r \,s\,\theta_\sigma\right)},
\ee
where $F_s$ are Klein factors, $a$ is a short-range cutoff, and fields
$\phi$ and $\theta$ obey commutation relations
\be
[\partial_x\phi_\alpha (x), \theta_{\alpha'}(x')] = - i \pi
\delta_{\alpha \alpha'}\delta(x-x').
\label{comm_rel}
\ee
Here $\alpha, \alpha' = \rho, \sigma$ label charge and spin degrees of
freedom, $\rho(x) = -(\sqrt{2}/\pi)\partial_x \theta_{\rho}$ and
$S_z(x) = -(\sqrt{2}/\pi)\partial_x \theta_{\sigma}$ are charge and
spin densities, and $J_c(x) = (\sqrt{2}/\pi)\partial_x \phi_{\rho}$ and
$J(x) = (\sqrt{2}/\pi)\partial_x \phi_{\sigma}$ are charge and spin
currents, respectively.

At length scales much longer than the Fermi wavelength, $1/k_F$, and
at low energy, the effective bosonized Hamiltonian takes the standard
form

\begin{align}
H = \sum_{\alpha = \rho,\s}{u_{\alpha} \over 2\pi} &\int dx \left\{\mc K_\alpha (\partial_x \phi_\alpha)^2 + \mc K_\alpha^{-1} (\partial_x \theta_\alpha)^2 \right\} \nn \\+&{\g \over {2\pi^2 a^2}}\int dx \cos \sqrt{8} \theta_\s + H_{\text{nl}} + H_{\text{so}}. \label{H_bos}
\end{align}

The parameters $u_{\s,\rho}$ correspond to the velocity of spin and
charge excitations, respectively, while $\mc K_{\s,\rho}$ are the
Luttinger parameters that characterize the sign and strength of
forward scattering interaction~\cite{Giamarchi}.

The relevance of the cosine term is controlled by the spin Luttinger parameter $K_\s$. When $K_\s < 1$, the cosine pins the spin-density variable, $\theta_\s$, leading to the formation of a spin gap \cite{LutherEmery}. On the other hand, when $K_\s > 1$, it is irrelevant and flows to $0$ under RG. In the latter case, its only effect is to renormalize the Luttinger parameters at low energies, and thus can be neglected.

As in a conventional Landau
theory\cite{Subir,Kardar}, classically, the transition to the ferromagnetic
state takes place when the coefficient in front of $S_z^2 \sim
(\partial_x \theta_\sigma)^2$ is tuned to be zero.
Close to the transition the coefficient at $S_z^2$ is small, corresponding to $\mc
K_\sigma \to \infty$. Consequently, the cosine term in Eq.~(\ref{H_bos}) is irrelevant, and thus neglected hereafter \cite{RuhmanKoziiFu}.

In contrast, there are higher-order terms that are typically  neglected near the LL fixed point, but, as we demonstrate below, become important at the ferromagnetic transition. These terms are taken into account in $H_{\text{nl}}$; they describe nonlinear couplings between spin density and spin current,
as well as between spin and charge sectors, that arise, for example, due to the finite curvature of the electronic dispersion. Since the coefficient of $S_z^2$ term vanishes at the transition, it
is necessary to keep these higher-order in
$S_z$ terms in order to stabilize the theory. Furthermore, as we will show, they are relevant in the RG sense, and thus play a qualitatively important role at low energies.

We further take advantage of spin-charge separation at the quadratic order and
completely neglect the charge sector, focusing only on the spin sector
controlling the transition to a state with finite magnetization
$S_z$\cite{noteChargeSector}. Then, the higher-order terms that are necessary for the stability of the theory are given by

\begin{align}
&H_{(3)} = \int dx \left\{  \beta_4\left[ S_z^4 + 6 S_z^2 J^2  \right] + \beta_2 (\partial_x S_z)^2  \right\} = \nn\\
 &\beta_0\int dx\left\{\left[\left( \partial_x \theta_\s \right)^4 +6\left( \partial_x \phi_\s \right)^2\left( \partial_x \theta_\s \right)^2\right] + 2 \left( \partial_x^2 \theta_\s \right)^2 \right\}, \label{Eq:H_3}
\end{align}
where $ \beta_0 = 4\beta_4 / \pi^4 = \beta_2 /\pi^2 =
({1/48\pi})(\partial_k^3 \ve_k )_{k = k_F}$, and we neglected terms that are less relevant. To derive Eq.~(\ref{Eq:H_3}), we used a standard point-splitting technique for normally-ordered operators, which allows, in principle, to calculate all higher-order terms. The details of such derivation can be found, for instance, in Ref.~\onlinecite{Teber}, and we thus do not present it here. The terms in
Eq.~(\ref{Eq:H_3}) describe the coupling between magnetization and spin
current, thus contributing to $H_{\text{nl}}$ in Eq.~(\ref{H_bos}). We
note that the sign of $\beta_0$ depends on the third derivative of the
dispersion near the Fermi energy. Here we focus on the more
interesting case when this
coefficient is positive, since the opposite case of $\beta_0 < 0$ leads to a
first-order transition, already at the Landau mean-field theory level.

We now examine the inversion-breaking contribution~(\ref{Eq:H_so}),
that is linear in spin current
$J=(\sqrt{2}/\pi)\partial_x\phi_\sigma$, and as expected induces a
finite spin current in the ground state. It can thus be absorbed into
(\ref{H_bos}) by shifting the spin current according to $\tilde J =
J + 2 \alpha_R k_F/\pi u_\s \mc K_\s$, with $\tilde J$ describing
fluctuations about the non-zero ground state spin current.

Focusing on the spin sector\cite{noteChargeSector}, and putting above
contributions together we obtain a low-energy Hamiltonian for the
one-dimensional itinerant Ising ferromagnetic transition,
\be
H =\int dx\left\{ \alpha_J \tilde J^2 + \alpha_S S_z^2+\beta_2 (\partial_x S_z)^2 - \lambda_3  \tilde J S_z^2+ \beta_4 S_z^4\right\}, \label{H_bos_2}
\ee
where $\lambda_3 ={6\pi\beta_2 \a_R k_F /u_\s \mc K_\s} $, $\alpha_J =
u_\s\mc K_\s \pi /4,$ and $\alpha_S = (u_\s \pi/ 4\mc K_\s) + (6
\alpha_R^2 k_F^2 \pi^2/u_\s \mc K_\s)$, and operators obey a
commutation relation
\be
[\tilde J(x), S_z(x')] = -i\frac{2}{\pi}\partial_x\delta(x-x').
\label{comm_rel}
\ee
The coefficient $\lambda_3$ is proportional to the spin current in
the system, and thus is a direct manifestation of inversion
breaking. This term is absent in Ref.~\onlinecite{Yang04}, where the
inversion-symmetric case has been considered.

We note that the above quantum Hamiltonian takes the form of a standard
Landau theory, but supplemented with a canonical commutation relation
(\ref{comm_rel}), with the spin current $\tilde J\sim \partial_x
\tilde \phi_\s,$ playing the role of the canonically conjugate
momentum density for the spin phase-field, $\theta_\sigma$.

The model~(\ref{H_bos_2}) has been derived for a one-dimensional metal with Rashba spin-orbit coupling. Note, however, that this model also applies to magnetic spin chains with Dzyaloshinskii-Moriya
interaction.

\section{Breakdown of quantum criticality at the FM transition}
 \label{Sec:FMtransition}

\subsection{Effective field theory}

To study the critical properties of the resulting model, we focus on
the partition function, $Z = \text{Tr}[\exp(-\beta H)]$, and express
it through the imaginary time functional integral over
commuting conjugate fields $\phi_\sigma,\theta_\sigma$  in a standard way

\be
Z = \int D\theta_\sigma D \Pi
\exp\left[- \int_0^{\beta} d\tau \int dx
\left( H - i \Pi \partial_\tau \theta_\sigma  \right)  \right],
\ee
where $\beta=1/T$ is the inverse temperature, and Hamiltonian $H$
depends on the canonically conjugate fields $\theta_\s(x)$
and $\Pi(x) = \partial_x \tilde \phi_\s(x)/\pi$. Integrating over the
momentum field $\partial_x \tilde \phi_\s$, we obtain $Z = \int
D\theta e^{-S}$, where the imaginary-time action for a quantum
itinerant PM-FM transition is given by

\begin{widetext}
\be
S = \int d^dx \, d\tau
\left\{\frac r2\left( \nabla \theta  \right)^2
+\frac K2\left( \nabla^2 \theta \right)^2+ \frac{B}2 \left( \partial_\tau \theta \right)^2 - \frac{i B_3}2 (\partial_\tau \theta) (\nabla \theta)^2 + \frac{B_4}8 \left[ (\nabla \theta)^2  \right]^2 \right\}. \label{Eq:action}
\ee
\end{widetext}
Above we dropped the index $\sigma$ for brevity, and generalized the
field theory to $d$ spatial dimensions, as it will be necessary for
the $\ve$-expansion analysis. Hereafter, we use $D=d+1$ for the total
number of space-time dimensions, while $d$ stands for a number of spatial
dimensions. We stress that the physically meaningful case corresponds to $d=1$ (quantum wire), while the extension of action (\ref{Eq:action}) to dimensions outside of 1d is used here as a mathematical tool only, in the spirit of $\ve$-expansion, to treat and control strong
critical quantum fluctuations.

In principle, all coefficients in the action ~(\ref{Eq:action}) can be
expressed through the parameters entering the microscopic Hamiltonian,
Eq.~(\ref{H_bos_2}). However, we prefer not to specify them explicitly,
treating them as phenomenological parameters, thereby emphasizing that
there is a number of microscopic contributions to this action,
beyond what we considered in the previous section.  The action
(\ref{Eq:action}) captures all universal properties of the itinerant
PM-FM transition in 1d, and, although in principle is derivable from
the microscopic model, can be written based purely on symmetry
arguments. We only require that $K, B, B_4 >0$; otherwise,
higher-order terms will be needed to stabilize the theory, and the
transition will be first-order even at the mean-field level.

As mentioned in the previous section, the first term in the
action~(\ref{Eq:action}), $ r(\nabla\theta)^2/2$, tunes the model to
the FM transition, at mean-field level $r<0$ ($r > 0$)
correspoinding to the ordered FM phase (disordered PM phase). The
third term in Eq.~(\ref{Eq:action}) describes fluctuations along
imaginary time, capturing the quantum nature of the transition.
Clearly, at the critical point, ($r=0$ in mean-field theory) higher-order terms in spatial gradient ($K$) and in the magnetization
($B_{3,4}$) are required to ensure the stability of the system to
large and nonuniform magnetization.  As we showed above, the term
proportional to $B_3$ arises from the coupling between magnetization
and spin current, i.e., proportional to $\lambda_3$ in the Hamiltonian
~(\ref{H_bos_2}). $B_3$ breaks inversion symmetry and will be of
special significance in our analysis.

We note that there is a certain similarity between our model
~(\ref{Eq:action}, \ref{H_bos_2}) and the compressible Ising model
\cite{BergmanHalperin1976}. In the latter case, the system can gain
energy by adjusting compressible lattice to the local spin
configuration. As a result, sufficiently close to the putative
critical point it is generically unstable to a discontinuous
development of a spontaneous magnetization (accompanied by a lattice
distortion), thereby undergoing a first-order transition. We
anticipate and indeed find a similar mechanism in our model. Namely, we expect that an
inversion-symmetry breaking that couples spin current and spin density,
$iB_3 (\partial_\tau \theta) (\partial_x \theta)^2$, will generically
drive the FM transition first-order associated with a discontinuous
jump in the spin current and magnetization.  There are, however, two
important differences between these two models. First, unlike the
compression modes of charge, the spin current is not a conserved
quantity. Second, the spin current, which is represented by the term
$i\partial_\tau\theta$, and the spin density, $\partial_x \theta$, are
not independent fields. Thus, a detailed analysis is required to which
we now turn.

\subsection{Harmonic fluctuations}

Away from the critical point, deep in the PM, $r > 0$ state,
higher-order gradients and nonlinearities are unimportant. In this
limit, the action (\ref{Eq:action}) reverts to that of a conventional
Luttinger liquid, described by a $1+1$ dimensional XY-model, with well
studied logarithmic phase correlations\cite{Giamarchi}.

At the critical point, $r=0$, within a harmonic approximation
(neglecting nonlinearities) the action maps onto that of a
well-studied $d+1$-dimensional smectic liquid
crystal\cite{Caille,deGennesProst,ChaikinLubensky}, with the imaginary
time axis and spin phase $\theta(\br,\tau)$ corresponding to the
smectic wavevector (layer normal) axis $\tau$ and the phonon
$u(\br,\tau)$, respectively.  At the critical point the fluctuations
are qualitatively enhanced, characterized by $z=2$ (rather than $z=1$)
dynamical exponent, with mean-squared fluctuations in the ground-state
($T=0$) given by
\bse
\begin{eqnarray}
\langle \theta^2\rangle_0
&=&\int^{a^{-1}}_{L^{-1}}\frac{d^dk d\omega}{(2\pi)^{d+1}}
\frac{1}
{B\omega^2 + K k^4}\ \ \ \ \\
&\approx&
\left\{\begin{array}{ll}
\frac{1}{2(2-d)\sqrt{B K}}C_{d}L^{2-d},& d < 2,\\
\frac{1}{4\pi\sqrt{B K}}\ln(L/a),& d = 2,\\
\end{array}\right.
\label{theta2}
\end{eqnarray}
\ese
where we defined a constant
$C_d=S_d/(2\pi)^d=2\pi^{d/2}/[(2\pi)^d\Gamma(d/2)]$, with $S_d$ a
surface area of a $d$-dimensional sphere ($S_1 = 2$, $S_2 = 2\pi$,
$S_3 = 4\pi$, etc.), and introduced a spatial infrared (IR) cutoff by
considering a system of finite spatial extent $L$ and a ultra-violet
(UV) cutoff $a$, set by the underlying lattice constant or a Fermi
wavelength $\sim k_F^{-1}$.  We note that for $d\leq 2$ and in
particular for the case of physical interest, $d=1$, harmonic quantum
fluctuations diverge (stronger than a conventional Luttinger liquid)
with system size, suggesting a qualitative importance of
nonlinearities.

The corresponding connected harmonic correlation function
\begin{equation}
C(\br,\tau)
=\langle\left[\theta(\br,\tau)-\theta({\bf 0},0)\right]^2\rangle_0\;
\label{C}
\end{equation}
is also straightforwardly worked out. At the critical point in $2+1$
space-time dimensions, in the ground-state ($T=0$) it is given by the
logarithmic Caill\'e form\cite{Caille}
\bse
\begin{align}
&C^{3D}(\br,\tau) = 2\int{d^2{k}d\omega
\over(2\pi)^3}{1-e^{i{\bf k}\cdot{\bf x}-i\omega\tau}\over K  k^4 + B
\omega^2}  \nonumber\\
 &= {1\over2\pi\sqrt{K  B}}
\left[\ln\left({x\over a}\right)-
\frac{1}{2}{\text{Ei}}\left({-x^2\over 4\lambda|\tau|}\right)\right],
\ \ \ \ \ \ \ \ \ \\
&\approx {1\over4\pi\sqrt{K  B}}\left\{\begin{array}{lcr}
\ln\left({x^2/ a^2}\right)&,&x\gg\sqrt{\lambda|\tau|} \;,\\ \ln\left({4\lambda |\tau|/ a^2}\right) &,&x\ll\sqrt{\lambda|\tau|}\;,\\
\end{array}\right.
\label{C3D}
\end{align}
\ese
where $\text{Ei}(x)$ is the exponential-integral function and $\lambda
= \sqrt{K/B}$.  As indicated in the last form, in the asymptotic
limits of $x\gg\sqrt{\lambda\tau}$ and $x\ll\sqrt{\lambda\tau}$ this
3D correlation function reduces to logarithmic growth with $x$
and $\tau$, respectively.

In the case $D=1+1$ of physical interest we instead
have\cite{TonerNelsonSm}
\bse
\begin{align}
&C^{2D}(x,\tau) = \int{d{k}d\omega\over(2\pi)^2}
{1-e^{i k x-i\omega\tau}\over K k^4 + B
\omega^2} \nonumber\\
 &= {1\over B}
\bigg[\left(\frac{|\tau|}{\pi\lambda}\right)^{1/2}
  e^{-x^2/(4\lambda|\tau|)} + \frac{|x|}{2\lambda}\mbox{erf}
\left(\frac{|x|}{\sqrt{4\lambda|\tau|}}\right)\bigg]\ \ \ \ \ \ \ \\
&\approx {1\over B}\left\{\begin{array}{lcr}
\left(|\tau|/\pi\lambda\right)^{1/2}&,
&x\ll\sqrt{\lambda|\tau|}\;,\ \ \ \ \\
|x|/2\lambda &, &x\gg\sqrt{\lambda|\tau|}\;,\ \ \ \ \\
\end{array}\right.
\label{Cuu2dT0}
\end{align}
\ese
where $\text{erf}(x)$ is the error function. Given these divergent
critical ground state fluctuations, it is important to examine the effect of
nonlinearities in the action (\ref{Eq:action}). We turn to this next.

\subsection{Perturbation theory and Ginzburg criterion}

To this end, it is helpful to first assess the role of nonlinearities
\begin{eqnarray}
S_{\text{nonlinear}}&=&\int d^d x d\tau
\left[-\oh i B_3 (\partial_\tau\theta)(\nabla\theta)^2
  + \frac{1}{8} B_4 (\nabla\theta)^4\right],\nonumber\\
\label{Snonlin}
\end{eqnarray}
using a conventional perturbation theory.
%

This can be done by computing perturbative corrections in
$S_{\text{nonlinear}}$ (\ref{Snonlin}) to any physical observable, e.g.,
the effective action itself. Following a standard field-theoretic
analysis, at low energies this can be encoded as corrections to the
couplings $B$ and $K$, with the leading contribution to $\delta B$,
summarized graphically in Eq.~(\ref{Eq:deltaB2}), and given by ($T=0$)
\begin{eqnarray}
\delta B&=&\oh B_3^2\int_{{\bf k},\omega} k^4
G({\bf k},\omega)^2\label{deltaBa} \nonumber\\
&=&\oh B_3^2\int{d^{d}k\over(2\pi)^{d}}\int_{-\infty}^{\infty}{d \omega\over2\pi}
{k^4\over(K k^4 + r k^2 + B\omega^2)^2}\label{deltaBb}\nonumber\\
&=&
\left[\frac{C_{d}\Gamma[(2-d)/2]\Gamma[(d+1)/2]}{8\pi^{1/2}}
{B_3^2\over (B K)^{3/2}}\xi^{2-d}\right]B\;,\label{deltaBc} \nonumber\\
&=&
\left[\frac{1}{8\pi}{B_3^2\over (B K)^{3/2}}\xi\right]B\;,
\;\text{for $d=1$}.\label{deltaBd}
\end{eqnarray}
\label{deltaB}
In above, we used $\theta({\bf k},\omega)$ two-point correlation
function, $G({\bf k},\omega)$ [see Eq.~(\ref{Eq:GreenFunction})], and focused on zero temperature ground state quantum fluctuations in $d < 2$, which allowed us to take the
UV-cutoff, $\Lambda\rightarrow\infty$. The dominant contribution from
the long-wavelength, low-energy modes is cutoff by the (Gaussian)
correlation length $\xi=\sqrt{K/r}$.

Since this nonlinear contribution grows with $\xi$, sufficiently close
to the critical point the correction $\delta B$ becomes comparable to
its bare microscopic value $B$. This signals a breakdown of the
harmonic theory near the critical point on length scales longer than
the Ginzburg scale
\begin{eqnarray}
\xi_{G}&=&
\left\{\begin{array}{ll}
\left[\frac{8\pi^{1/2}}{C_{d}\Gamma[(2-d)/2]\Gamma[(d+1)/2]}
{(B K)^{3/2}\over B_3^2}\right]^{1/(2-d)},&  d < 2,\nonumber\\
\frac{8\pi (B K)^{3/2}}{B_3^2},& d = 1,
\end{array}\right.\nonumber\\
\label{xiG}
\end{eqnarray}
defined by the value of $\xi$ at which $|\delta
B(\xi_{G})|=B$. Equivalently, this also gives the Ginzburg criterion
$r_G = K\xi_G^{-2}$, corresponding to a ``distance'' to the critical point
at which critical fluctuations qualitatively modify the predictions of
the harmonic analysis at the Gaussian fixed point.

\subsection{RG analysis and $\ve$-expansion}

To describe the critical properties beyond the Ginzburg scale,
$\xi_G$, near the critical point with $|r| < r_G$ -- i.e., to make sense
of the IR divergent perturbation theory found in
Eq.~(\ref{deltaBd})\ -- requires a renormalization group analysis. As we
discuss in Sec.~\ref{Sec:smectics} this was first done at the critical dimension of
$d=2$ in the context of a smectic-A to smectic-C liquid crystal phase
transition in a seminal work by Grinstein and Pelcovits
(GP)\cite{GrinsteinPelcovits81,GrinsteinPelcovits82}.

To this end, we employ the standard momentum-shell RG transformation
\cite{WilsonRG} by separating the field into long and short scale
contributions according to $\theta(\br,\tau)=\theta_{<}(\br,\tau) +
\theta_{>}(\br,\tau)$ and perturbatively in nonlinearities,
$S_{\text{nonlinear}},$ integrate out the short-scale (high momenta)
fields $\theta_{>}(\br,\tau)$, that take support inside an
infinitesimal cylindrical momentum-frequency shell $\Lambda
e^{-\delta\ell} < k_> < \Lambda\equiv 1/a$,
$-\infty<\omega<\infty$. Purely for convenience, we follow this with a
rescaling of lengths, times and the long wavelength part of the field
in real space:

\be
\bx = e^{\delta\ell} \bx', \qquad \tau = e^{z\delta\ell} \tau',
\qquad \theta_<(\bx, \tau) = e^{\chi\delta\ell}\theta'(\bx',\tau'),
\label{rescaleRG}
\ee
so as to restore the UV cutoff $e^{-\delta\ell}\Lambda$ back to
$\Lambda=1/a$. Above, $z$ is a dynamical exponent, $\chi$ is a field
dimension, and $\ell$ is ``RG time''.

The above rescaling leads to zeroth-order RG flows of the effective
couplings after coarse-graining by a factor $e^\ell$
\begin{eqnarray}
r(\ell)&=&e^{(d+z-2+2\chi)\ell}r,\nonumber \\
\label{flow_r}
K(\ell)&=&e^{(d+z-4+2\chi)\ell}K,\nonumber \\
\label{flow_K}
B(\ell)&=&e^{(d-z+2\chi)\ell}B,\nonumber \\
\label{flow_B}
B_3(\ell)&=&e^{(d-2+3\chi)\ell}B_3,\nonumber \\
\label{flow_B3}
B_4(\ell)&=&e^{(d+z-4+4\chi)\ell}B_4.
\label{flow_B4}
\label{flow0}
\end{eqnarray}

To assess the importance of nonlinearities relative to harmonic terms,
it is convenient (but not necessary) to keep the quadratic terms fixed
under the RG flow, i.e., to choose $K(\ell)=K$ and $B(\ell)=B$,
corresponding to a choice of $z=2$, $\chi = (2-d)/2$. With this, we
find $B_3(\ell)=e^{\oh(2-d)\ell}B_3$, $B_4(\ell)=e^{(2-d)\ell}B_4$,
reflecting their importance at the critical point $r=0$, below the
upper-critical dimension $d_c = 2$. This is consistent with our
finding in Eq.~(\ref{deltaBd}) of a divergent perturbation theory for $d <
2$. Since the nonlinearites are irrelevant for $d > d_c$, and thus are
only weakly relevant just below $d=2$, we expect to control our
perturbative RG analysis for $d < 2$ by an $\ve$-expansion in $\ve =
2-d$. As discovered by Wilson and Fisher in the context of classical
ferromagnet\cite{WFepsilon}, this gives us a
controlled method to analyze the critical properties of a physical
$d=1$ ferromagnetic wire, by extrapolating via $\ve = 1$.

The leading one-loop order RG comes from integrating out the
high-momentum modes, $\theta_{>}(\br,\tau),$ perturbatively in
$S_{\text{nonlinear}}$. The contributions are of the same form as in a
direct perturbation theory (e.g., $\delta B$ in (\ref{deltaB})), but
with the correction kept small by the infinitesimal momentum shell,
$\delta\ell$. Relegating the technical details to Appendix \ref{app:diagrams}, the
result of this coarse-graining RG procedure is encoded in $\ell$
dependent couplings, that we find to satisfy the flow equations (we focus on the $T=0$ case):
\begin{widetext}
\begin{eqnarray}
\frac{d B}{d\ell} & = & \left( d - z + 2\chi\right)B +
\frac{B_3^2 \gamma_d}{B^{1/2} (K+\tr)^{3/2}}, \nonumber   \\
\frac{dB_3}{d\ell} & = & \left( d-2  + 3\chi  \right)B_3 -
\frac{B_3\gamma_d}{ B^{3/2} (K+\tr)^{3/2} d}\left[\left(d+ 2\right)
  B B_4 +  2B_3^2\right], \nonumber \\
\frac{dB_4}{d\ell} & = & \left( d-4 + z + 4\chi \right) B_4
-\frac{\gamma_d}{B^{5/2} (K+\tr)^{3/2} (d^2+2d)}\left[(d^2+6d+20)B^2
  B_4^2 + 4(d+8)B B_4B_3^2 + 12 B_3^4 \right], \nonumber \\
\frac{dK}{d\ell} & = & \left( d-4 + z + 2\chi \right)K
- \frac{B_3^2  \gamma_d}{2 B^{3/2}(K+\tr)^{5/2} d(d+2)}
\left[  (2K+\tr)(K+\tr)(d+2) -3\tr^2 \right], \nonumber \\
\frac{d\tr}{d\ell} & = &  \left( d-2 + z + 2\chi \right)\tr
+ \frac{2 \gamma_d}{B^{3/2} (K+\tr)^{1/2}}\left(B B_4 + B_3^2 \right)
\left( 1 + 2/d \right),
\label{Eq:RGset}
\end{eqnarray}
\end{widetext}
where we defined $\gamma_d = C_d\Lambda^{d-2}/8$ and $\tilde r\equiv
r/\Lambda^2$.

As discussed in detail in Appendix \ref{app:diagrams}, we note that coarse-graining
also generates a term $i\delta\alpha_R\partial_{\tau} \theta$, that is a
correction to the average ground state spin current, that thus flows
under RG (much like an order parameter in an ordered state). This
operator can be shifted away by redefining the average spin current,
corresponding to a shift in $\theta$ by
$\propto\delta\alpha_R\tau$. From the $B_3$ operator this then
generates a correction to $(\nabla \theta)^2$, i.e., a
$\delta\alpha_R$ correction to the critical coupling $r$, which (along
with two other contributions, $B_3^2$ and $B_4$ tadpole) has been
included in the last equation in (\ref{Eq:RGset}). Anticipating the
connection of the quantum FM transition with the classical smectic
liquid crystal, we note that this procedure is analogous to the RG
flow of the smectic ordering wave vector, as discussed in Ref.~\onlinecite{GrinsteinPelcovits81}.

To bring out the physical content of the above flow equations and to
simplify the mathematical analysis, it is convenient to use
(\ref{Eq:RGset}) to construct a flow of two dimensionless couplings

\be
g_1 = \frac{B_3^2}{(B K)^{3/2}}\gamma_d,
\qquad g_2= \frac{B_4}{(B K^3)^{1/2}}\gamma_d,
\label{Eq:g1g2}
\ee
where for consistency of the $\ve$-expansion $d$ must be evaluated at the
upper critical dimension, i.e., $\gamma_d \to \gamma_2 = 1/16\pi$.

These couplings can be shown to satisfy dimensionless RG flow
equations
\bse
\label{Eq:RGg1g2}
\begin{eqnarray}
\frac{dg_1}{d\ell} &=& \ve g_1 - g_1 \left(\frac{11 g_1}4
 + 4g_2\right), \\
\frac{dg_2}{d\ell} &=& \ve g_2 - g_2 \left( \frac{19 g_1}4
+ \frac{9 g_2}2 \right) - \frac{3 g_1^2}2,
\end{eqnarray}
\ese
which we note are independent of the arbitrary rescaling exponent $z$
and $\chi$ (that only acquire physical content if $B$ and $K$ are
chosen to be kept fixed under coarse-graining). In the above, for
consistency of the $\ve$-expansion we also evaluated $d$ at $d_c$,
i.e., set $d=2$ in the quadratic terms on the right hand side, and,
focusing on the vicinity of a critical point, set $r=0$.

In terms of $g_1$ and $g_2$ the flow of the harmonic couplings is then
given by
\begin{eqnarray}
\frac{d B}{d\ell} & = & \left[d - z + 2\chi + g_1\right]B \nonumber \\
& = & \left[d - z + 2\chi - \eta_B\right]B,\nonumber\\
\frac{dK}{d\ell} & = & \left[d-4 + z + 2\chi - \oh g_1\right] K \nonumber \\
& = & \left[d - 4 + z + 2\chi + \eta_K\right]K,\nonumber\\
\frac{d \tr}{d\ell} & = & \left[d-2 + z + 2\chi - 2(g_1 + g_2)\right]\tr \nonumber
\\ & + & 4(g_1 + g_2)K,
\label{Eq:RGdimensionless}
\end{eqnarray}
where we implicitly defined the anomalous exponents $\eta_{B,K}$, that
flow to universal values at a critical point $g_1^*, g_2^*$. The
last term in the $\tr$ equation corresponds to the fluctuation-driven
downward shift of the critical point $\tr_c$.  The dynamical exponent
$z$, defined by the relation $\tau_\xi\sim\xi^z$ (see (\ref{rescaleRG}))
between the correlation
time $\tau_\xi$ and correlation length $\xi\sim \tilde r^{-\nu}$, and the
correlation length exponent $\nu = 1/y_r$ (inverse of the eigenvalue
$y_r$ of $\tr$) are then determined by
\begin{eqnarray}
 z  &   = &  2 - \oh(\eta_B + \eta_K) = 2 + \frac{3}{4}g_1^*,\nonumber\\
\nu & = &\left(2 - \frac{3}{2} g_1^* - 2g_2^*\right)^{-1}.
\label{Eq:z_nu_general}
\end{eqnarray}
\begin{figure}
\includegraphics[width=\columnwidth]{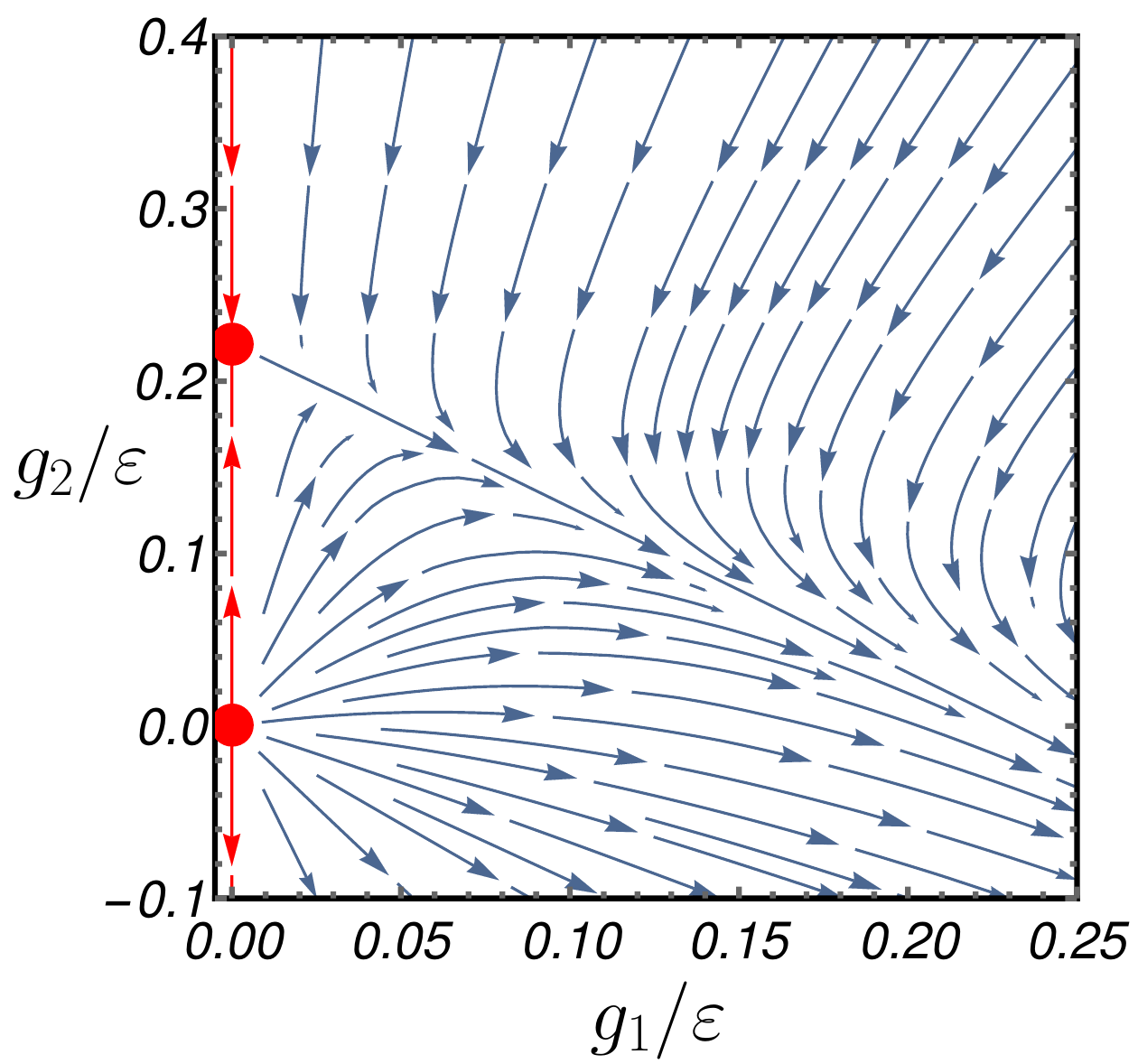}
\caption{RG flow of the parameters $g_1$ and $g_2$ defined in Eq.~(\ref{Eq:g1g2}) describing the FM transition. The flow is given by Eq.~(\ref{Eq:RGg1g2}). Two unstable fixed points are located at the line $g_1=0$, see Table~\ref{Table1}. We see that, at large enough RG time $\ell$, $g_2$ flows to negative values, thus necessarily resulting in a first-order transition.}
\label{Fig:RGFM}
\end{figure}
\begin{figure}
\includegraphics[width=8.25cm]{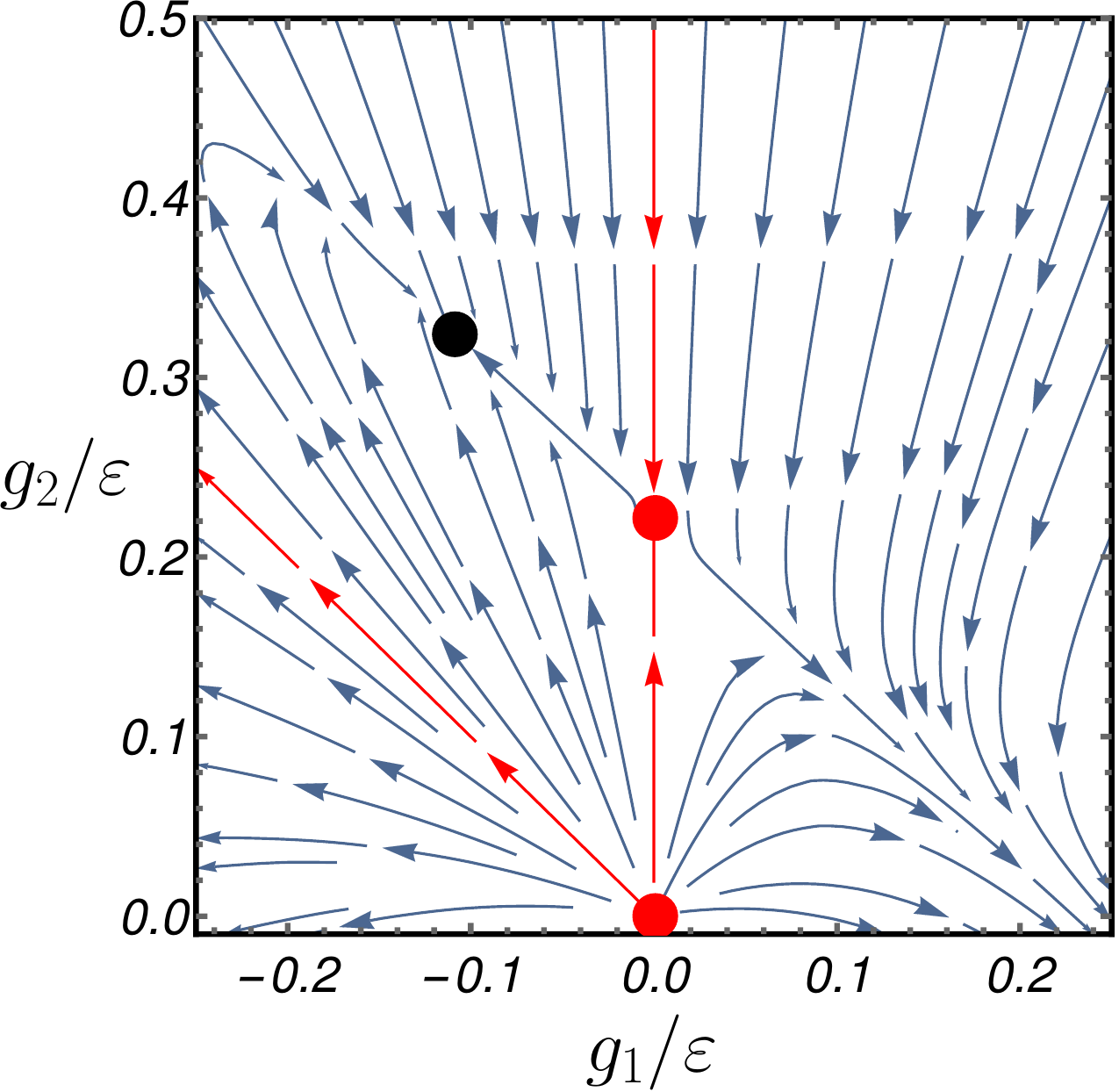}
\caption{Global phase diagram of the RG flow described by Eq.~(\ref{Eq:RGg1g2}) for both the FM and SmA-C transitions. The region $g_1>0$ describes a runaway flow signaling a first-order phase transition. The line $g_1=0$ governs the FM transition in the inversion-symmetric system, which is controlled by the IS fixed point. The region $0<-g_1<g_2$ is controlled by an interacting fixed point (black) and describes the second-order SmA-C transition in a magnetic field. The line $-g_1 = g_2$ describes the SmA 'critical phase', with the parameters $g_1$, $g_2$ flowing to the Sm fixed point (not shown in this figure, see also Fig.~\ref{Fig:RGSm}). }
\label{Fig:RGboth}
\end{figure}

The flow diagrams corresponding to (\ref{Eq:RGg1g2}) are shown in
Figs.~\ref{Fig:RGFM}, \ref{Fig:RGboth}, and \ref{Fig:RGSm}.  As anticipated based on the perturbative analysis and
power counting, for $d < 2$, in the presence of quantum fluctuations
the Gaussian (G) critical point is unstable to interactions. Simple
analysis shows that there are three non-Gaussian critical points: (i)
inversion-symmetric (IS) with $g_1=0, g_2 > 0$, (ii) smectic (Sm) with
$g_1 < 0, g_2 = |g_1|$, (iii) smectic-A to C transition (SmAC) with
$g_1 < 0, g_2 \neq |g_1|$, summarized in Table \ref{Table1}.
We next study the physical significance of these critical points and
their critical properties, noting that $g_1>0$ and $g_1<0$, respectively, are realizable in a FM wire (studied next) and the SmA-C liquid crystal (studied in Sec.~\ref{Sec:smectics}). We emphasize that all (non-Gaussian) fixed points find realization in certain physical systems, see Secs.~\ref{Sec:IS}, \ref{Sec:smectics}.

\subsection{Ferromagnetic transition}

It is clear from our derivation of the ferromagnetic model,
Eq.~(\ref{Eq:action}), and the definition of $g_1$, Eq.~(\ref{Eq:g1g2}), that for a FM
wire the case of physical interest is $g_1 \ge 0$;  the other half
plane, $g_1 < 0$, does not appear to be accessible to the FM system. However, as we discuss in Sec.~\ref{Sec:smectics}, it does find a physical realization in smectic liquid crystals.

\subsubsection{Inversion-symmetric FM \label{Sec:IS}}

The inversion-{\em symmetric} FM is constrained by $B_3 = g_1 = 0$. In this subspace, the flow for $g_2$ reduces to (see Eqs.~(\ref{Eq:RGg1g2})-(\ref{Eq:RGdimensionless}))

\beq
\frac{dg_2}{d\ell} &=& \ve g_2 - \frac{9}{2} g_2^2,\label{g2flowIS}
\eeq
giving the nontrivial IS critical point,

\be
g^*_1 = 0, \qquad g^*_2 = \frac{2\ve}9, \qquad r^* =
-\frac{4\ve\Lambda^2}9,
\ee
previously studied by Kun Yang\cite{Yang04}. This fixed point
controls an inversion-symmetric, itinerant PM-to-FM quantum phase
transition, and to one-loop order is characterized by
\begin{align}
&\eta_B=\eta_K = 0,\qquad z = 2, \nonumber \\ &\nu = \frac{1}{2 -
  4\ve/9} \approx \oh \left(1 + \frac{2\ve}9 \right) \approx \frac{11}{18},
\label{Eq:z_nu_IS}
\end{align}
where the last expression for the correlation length exponent $\nu$ was evaluated for the physical
case of one-dimensional FM wire, $d=1$ ($\ve = 1$). The other critical exponents, up to linear order in $\ve$, are given by

\begin{align}
&\eta = 4 - d - z - 2\chi \approx 0, \nonumber \\
&\gamma = (2-\eta) \nu \approx \left(1+ \frac{2\ve}9 \right) \approx \frac{11}9, \nonumber \\ &\beta = \frac{\nu}2 (d + z + \eta - 2) \approx \frac12 \left( 1- \frac{5\ve}{18}  \right) \approx  \frac{13}{36},
\end{align}
where $\eta$ is an anomalous dimension, $\gamma$ is susceptibility exponent, and $\beta$ is magnetization exponent \cite{commentKunYang}.

\subsubsection{Inversion-asymmetric FM}

We now turn to the main focus of the paper, namely the FM phase
transition in an inversion-{\em asymmetric} itinerant ferromagnet, with
$B_3\neq 0$.

Our key observation is that the inversion-symmetric $g_1=0$ fixed
point discussed in the previous subsection is unstable to $g_1\neq 0$, with the
symmetry-breaking growth characterized by

\be
\frac{d g_1}{ dl} = \frac{\ve g_1}9.
\ee

It is clear from the RG flows (see Fig.~\ref{Fig:RGFM}), that there is no
stable critical point for $g_1 > 0$. Similar fluctuation-driven
runaway flows have been discussed in the literature, most prominently in the context of a normal-to-superconductor and (mathematically
related) nematic-to-smectic-A phase
transitions~\cite{Halperin74,ChenLubenskyNelson78}. For small $\ve$,
the absence of a stable fixed point was demonstrated via a detailed RG
analysis to be a signature of a fluctuation-driven first-order
transition\cite{commentHLM}. Other examples include crystal-symmetry
breaking fields in $O(N)$ magnets\cite{RudnickNelson} and
isotropic-to-tetrahedratic phase transition\cite{RadzihovskyLubenskyT4}.

Generically, to demonstrate a fluctuation-driven first-order
transition requires a detailed RG computation of the free
energy\cite{RudnickNelson}. Here, instead we argue that the
inversion-asymmetric itinerant PM-FM phase transition is driven
first-order based on qualitative arguments, leaving a detailed
computation of the free energy to future studies.

To this end we first observe (see Fig.~\ref{Fig:RGFM}), that for a given bare $B_4 > 0$ and non-zero $B_3$, sufficiently close to the $g_1=0$ critical point, quantum fluctuations with $g_1 > 0$ always drive $B_4$ negative. RG analysis allows us to map a nearly critical strongly fluctuating system at small $r$ to a coarse-grained noncritical system at large $r(\ell_*) \sim \Lambda^2$. Then, to find a transition, we can simply minimize the {\it coarse-grained} Hamiltonian density that approximates the ground-state energy density ${\cal E}_{gs}(\tilde J,S,r)$

\be
{\cal  E}_{gs}= \frac{1}{2B}\tilde J^2 - \frac{B_3}{2B} \tilde J S^2 + \frac r 2 S^2 +
\left(-\frac{|B_4|}8 + \frac{B_3^2}{8B}\right) S^4 + B_6 S^6,
\label{Eq:Heff}
\ee
over $\tilde J$ and $S$, where $\tilde J$ is proportional to the fluctuations of the spin current about its average value and $S$ is proportional to magnetization.  Minimizing
over $\tilde J$ gives a standard quartic form with the renormalized $B_4$ driven negative and $B_6$ included for the overall stability

\be
{\cal  E}_{gs} =  {r(\ell_*)\over 2} S^2 - \frac{|B_4(\ell_*)|}8 S^4 + B_6(\ell_*) S^6. \label{Eq:Heff1}
\ee
We emphasize that all couplings $r(\ell_*)$, $B_3(\ell_*)$, $B_4(\ell_*)$, $B_6(\ell_*)$ in Eqs.~(\ref{Eq:Heff})-(\ref{Eq:Heff1}) are solutions to the RG flow equations (\ref{Eq:RGset}) evaluated at $\ell_*$ defined by $r(\ell_*) \sim \Lambda^2,$ or, equivalently, $\ell_*(r_*) = \ln[\xi(r_*)/a]$.

Because the mapped system lies outside the Ginzburg region, where
fluctuations are small, its ground state state energy and the associated
transition can be computed within mean-field approximation. We thus find
that a weakly first-order transition takes place at
$r_*$, with a magnetization jump $S_*$ implicitly determined by

\bse
\beq
r_* &\approx& \frac{B_4^2(\ell_*)}{128 B_6(\ell_*) e^{2\ell_*}},\\
S_* &\approx& \left(\frac{8 r_*}{| B_4(\ell_*)|}\right)^{1/2}.
\eeq
\ese

To summarize, we find a strong,
general result, that the itinerant PM-FM
transition in the absence of inversion symmetry {\em must be first order}. This
contrasts qualitatively from the inversion-symmetric case and the Hertz-Millis expectation
of a continuous transition.

\section{Smectic-A -- C transition in a magnetic field
\label{Sec:smectics}}

As we discuss in detail below, quite remarkably the model of the
itinerant PM-FM quantum phase transition studied above is
mathematically equivalent to that of a classical $D$-dimensional
smectic-A to smectic-C liquid crystal transition in a magnetic
field.\cite{deGennesProst,ChaikinLubensky} We recall that a smectic-A
liquid crystal is a one-dimension density wave (a one-dimensional
periodic array of 2D liquid sheets) of rod-like constituents
(calamitic molecules) defined by director $\hat n$ aligned along the
smectic layer normal, $\hat\kappa$, spontaneously breaking rotational
and one-dimensional translational symmetries. To match the notation of
the FM action (\ref{Eq:action}), we denote the latter spatial axis (conventionally denoted by
$z$) to be $\tau$, with the smectic classical Hamiltonian given by

\begin{widetext}
\be
H_{\text{SmA}} = \int d^2x d\tau\left[\frac{K}{2}(\nabla_\perp^2 u)^2
+ \frac{B}{2}\left(\partial_\tau u - \oh \left(\nabla_\perp u\right)^2\right)^2\right],
\label{Hsm}
\ee
\end{widetext}
where $u$ is the smectic scalar phonon field that describes distortions of smectic layers along the layer normal. We note that the underlying rotational invariance of the smectic phase
is encoded through the nonlinearities appearing only via a fully rotationally
invariant strain, $u_{\tau\tau} = \partial_\tau u - \oh (\nabla_\perp u)^2$,
that at harmonic level reduces to the absence of the quadratic
$(\nabla_\perp u)^2$ term in Eq.~(\ref{Hsm}). The latter would otherwise incorrectly penalize
a rotation of the smectic layers by an infinitesimal angle
$\theta\approx\nabla_\perp u$\cite{GrinsteinPelcovits81,ChaikinLubensky}.

\begin{figure}[b]
\includegraphics[width=\columnwidth]{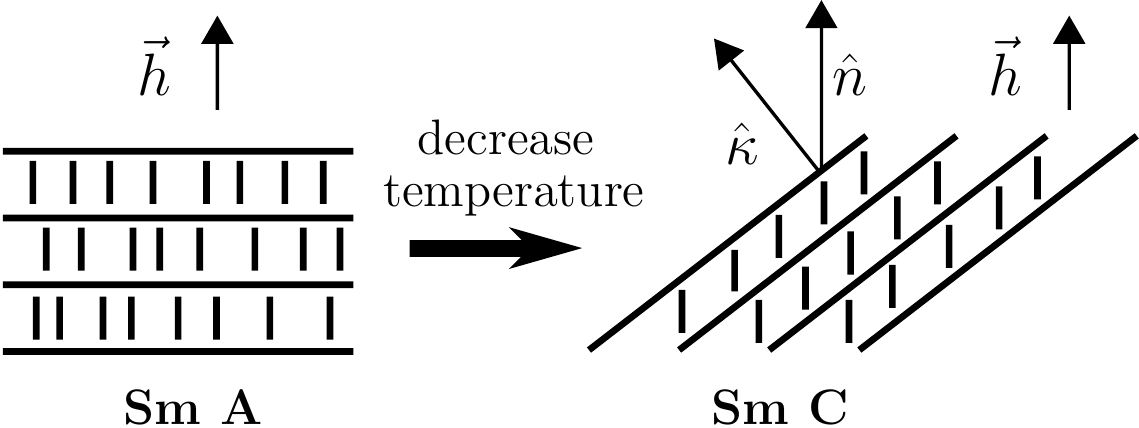}
\caption{ Schematic representation of the smectic-A to smectic-C transition in a magnetic field, $\vec h$. The transition occurs via tilting of the layers, and is characterized by a non-zero angle between the molecules' director $\hat n$ and the smectic layer normal, $\hat \kappa$ (see Ref.~\onlinecite{GrinsteinPelcovits82}).
}
\label{Fig:smecticC}
\end{figure}

As illustrated in Fig.~\ref{Fig:smecticC}, a smectic-C
liquid crystal is distinguished by a spontaneous molecular tilt of
$\hat n$ into the smectic planes, i.e., $\hat n \cdot \hat\kappa < 1$.
The associated XY order parameter is the $\vec c$-director, a
projection of $\hat n$ into the smectic planes, characterized by an
effective Hamiltonian (within a single elastic constant $K$ approximation)
\be
H_{\text c} = \int d^2x d\tau\left[\frac{K}{2}(\nabla_\perp\vec c)^2
+ \frac{r}{2} c^2 + \frac{\lambda}{4} c^4\right],
\label{Hac}
\ee
with the reduced temperature $r\sim T-T_{AC}$ driving the
AC transition.

A model for smectic A-C transition was proposed by Chen and Lubensky \cite{ChenLubensky76}, and then extensively studied by Grinstein and
Pelcovits\cite{GrinsteinPelcovits82}, who demonstrated that, despite the nontrivial
coupling of the $\vec c$-director XY model (\ref{Hac}) to the smectic
phonon $u$ elasticity (\ref{Hsm}), the criticality remains of a
conventional XY-model (superfluid Helium-4 transition) universality
class, as originally conjectured by de Gennes\cite{deGennesProst}.

In a magnetic field $\vec h$, the liquid crystal molecules with positive
diamagnetic anisotropy align along the field, with the energetics
governed by

\bse
\label{Hfield}
\beq
H_{\text{field}} &=& -\oh \chi_a\int d^2x d\tau (\hat n\cdot\vec h)^2,\\
 &\approx& -\frac{h^2}2 \chi_a\int d^2x d\tau
\left[1 - (\vec c - \nabla_\perp u)^2\right]
\eeq
\ese
with $\chi_a$ the associated susceptibility. As was discussed by GP
and illustrated in Fig.~\ref{Fig:smecticC}, because the molecules are
locked along the field, $\vec h$, the AC transition in a magnetic
field is associated with the spontaneous tilt of smectic layers toward the
magnetic field axis.

Furthermore, because the underlying rotational symmetry is broken by
the $\vec h$ field, the nature of transition is qualitatively
modified. Indeed, neither Sm-A nor Sm-C is any longer rotationally
invariant. It is clear that the magnetic field locks the $\vec
c$-director to layer tilt, $\vec c \approx \nabla_\perp u$, allowing
one to reexpress $H_c[\vec c\rightarrow\nabla_\perp u]$ in terms of
$u$, formally done by integrating out the $\vec c$-director. Not
surprisingly, this leads to a smectic-like Hamiltonian, but with the
rotational symmetry broken by the magnetic field $\vec h$, and the
smectic A-C transition described by a classical Hamiltonian\cite{GrinsteinPelcovits82}
\begin{widetext}
\be
H_{\text{SmACfield}}= \int d^{D-1}x d\tau
\left[\frac K2(\nabla^2_{\perp}u)^2
+ \frac r2(\nabla_{\perp} u)^2 + \frac{B}2(\partial_\tau u)^2
- \frac{B_3}2 (\partial_\tau u) (\nabla_{\perp} u)^2
+ \frac{B_4}8 (\nabla_{\perp} u)^4\right],
\label{Eq:HsmACfield}
\ee
\end{widetext}
generalized to $D=d+1$ dimensions. Clearly, at the Sm critical point, which is characterized by $r=0$
and by a special relation of the nonlinearities, $B_3 = B_4 = B$,
this reduces to the fully rotationally invariant smectic elasticity, Eq.~(\ref{Hsm}).

\begin{figure}[h!]
\includegraphics[width=\columnwidth]{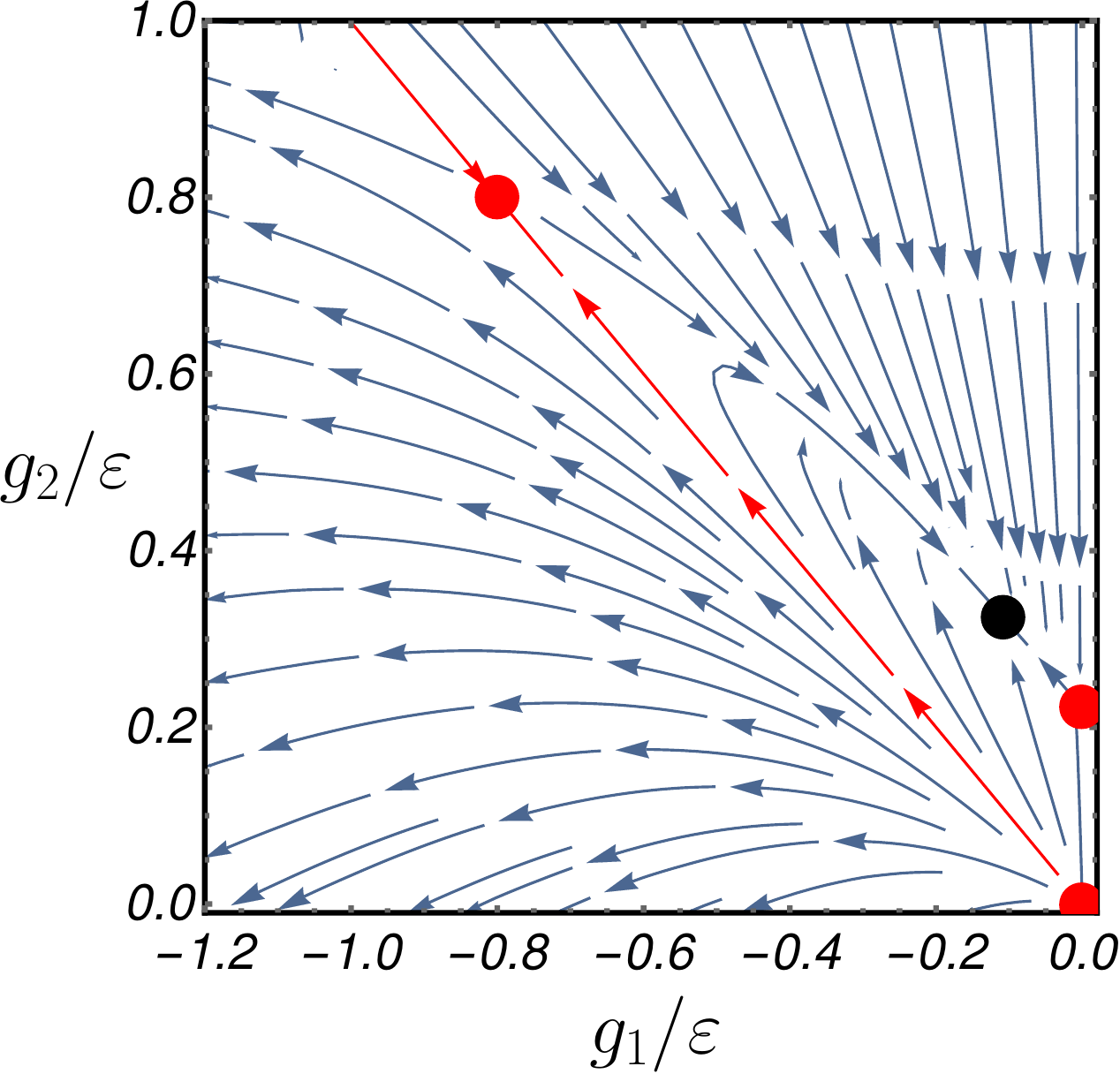}
\caption{ The RG flow for the SmA-C transition in a magnetic field described by Eq.~(\ref{Eq:RGg1g2}) in the region $g_1<0$ ($g_1<0$ part of Fig.~\ref{Fig:RGboth}). A stable fixed point (black) controls the second-order transition. The region $g_2<|g_1|$ corresponds to the mechanical instability of the system, and, thus, describes a first-order transition. The separatrix $|g_1|=g_2$, which separates two regions (red line), corresponds to the SmA line, and is controlled by the Sm fixed point.
}
\label{Fig:RGSm}
\end{figure}

We next note that this classical Hamiltonian for the smectic A-C
transition is identical in form to that of the itinerant quantum PM-FM
transition (\ref{Eq:action}), studied in the previous section. However, a key
difference is the absence of $i$ in the $B_3$ term of the classical
problem. Thus, the smectic A-C transition maps directly onto our FM
transition analysis in Sec.~\ref{Sec:FMtransition}, but with the dimensionless coupling
$g_1\sim -B_3^2 < 0$. It thus allows us to access the $g_1 < 0$ half of
the flow diagram in Fig.~\ref{Fig:RGboth}, and in particular the two additional
critical points, that with some foresight we earlier denoted by Sm and
SmAC (see also Table~\ref{Table1} and Fig.~\ref{Fig:RGSm}).

\begin{center}
\begin{table*}[t]
\begin{tabularx}{1.\textwidth}{|l| l | l | l | l | l | l| l| l| l | l | X |}
    \hline Fixed point & $g_1$ & $g_2$ & $\eta_B$ & $\eta_K$ & $r$ & $z$ & $\chi$ & $\nu$ & $\gamma$ & $\beta$ &\text{stability} \\ \hline
    G & 0 & 0 & 0 & 0 & 0 & 2 & $\ve/2$ & 1/2 & 1 & 1/2 & unstable \\ \hline
 IS & 0 & $2\ve/9$ & 0 & 0 & $-4\ve/9$ & 2 & $\ve/2$ & $(9 + 2\ve)/18 $ & $1+ (2\ve/9) $ & $(18-5\ve)/36$ & unstable  \\ \hline
 Sm & $-4\ve/5$ & $4\ve/5$ & $4\ve/5$ & $2\ve/5$  &0 & $2-(3\ve/5)$ & $3\ve/5$ & -- & -- & -- & unstable\\
    \hline SmAC & $-4\ve/37$ & $12\ve/37$ & $4\ve/37$ &  $2\ve/37$ & $-16\ve/37$ & $2-(3\ve/37)$ & $19\ve/37$ & $(37+9\ve)/74$ & $1+(\ve/74)$ & $(37-10\ve)/74$ & stable \\ \hline
\end{tabularx}
\caption{The fixed points of the one-loop RG flow (\ref{Eq:RGg1g2}) describing the FM and SmA-C transitions. In the corresponding regions of stability, critical points are characterized by an anomalous dimension $\eta$, correlation length critical exponent $\nu$, susceptibility exponent $\gamma$, and order parameter exponent $\beta$. }
\label{Table1}
\end{table*}
\end{center}

The most ubiquitous case of a three-dimensional smectic
lies right at the upper critical dimension, $D= D_{cr} = 3$ ($d=2$),
and has been extensively analyzed in Refs.~\onlinecite{GrinsteinPelcovits81,GrinsteinPelcovits82}.  Our
results, summarized by equations
(\ref{Eq:RGg1g2})-(\ref{Eq:RGdimensionless}) are a generalization of
GP's work to arbitrary dimension and in particular to $D < 3$, where
the nonlinearities are relevant (rather than marginally
irrelevant\cite{GrinsteinPelcovits81,GrinsteinPelcovits82}) and lead to nontrivial fixed points,
illustrated in Fig.~\ref{Fig:RGSm}.

As is clear from the RG flows, the Gaussian and IS critical points,
discussed in the context of the FM transition, are unstable to the Sm and SmAC
fixed points. The SmAC critical point is the one with global stability
(to order $\ve$) and thus controls the smectics-A to smectic-C phase
transition in a magnetic field. It is given by


\be
g_2^* = -3g_1^* =\frac{12\ve}{37},  \qquad r^* = -\frac{16\ve}{37},
\ee
and is characterized by the anomalous universal exponents

\begin{align}
&\eta_B = 2\eta_K = \frac{4\ve}{37} \approx \frac{4}{37}, \nonumber \\
&z \approx 2-\frac{3\ve}{37} \approx \frac{71}{37}, \nonumber \\ &\eta \approx \frac{17\ve}{37} \approx \frac{17}{37}, \nonumber \\  &\nu \approx \frac12\left( 1 + \frac{9\ve}{37}  \right) \approx \frac{23}{37}, \nonumber \\ &\gamma \approx 1 + \frac{\ve}{74} \approx \frac{75}{74}, \nonumber \\ &\beta \approx \frac12\left( 1 - \frac{10\ve}{37}  \right) \approx \frac{27}{74},
\end{align}
evaluated for the only case of physical
interest, the two-dimensional smectic, $D=2$ ($\ve=1$).

From the global phase diagram perspective (see Fig.~\ref{Fig:RGboth}), we thus find that the phase
transition is continuous for $g_1 < 0$ (controlled by the SmAC
critical point in the region of mechanical stability, $g_2>|g_1|$; in the region $g_2<|g_1|$, higher-order terms are needed to stabilize a theory, and a transition is automatically first-order) and is driven by fluctuations to be first-order for
$g_1 > 0$. The two regimes are then separated by the
inversion-symmetric tricritical IS point at $g_1=0$.

The other critical point is the unstable Sm fixed point, characterized
by $-g_1=g_2 > 0$ or, equivalently, $B_3^2 = B_2 B_4$. We note that in
this coupling subspace, the nonlinearities assemble into a complete
square of a nonlinear strain tensor
\begin{align}
\frac{B}2 (\partial_\tau u)^2
- \frac{B_3}2 (\partial_\tau u) (\nabla_\perp u)^2
+ \frac{B_4}8 (\nabla_\perp u)^4\nonumber \\
= \frac{B}2 \left[\partial_\tau u
 - \frac12 \sqrt{\frac{B_4}{B}} (\nabla_\perp u)^2\right]^2,
\end{align}
that after an inconsequential rescaling of the phonon $u$ reduces to a
fully rotationally invariant nonlinear smectic elasticity,
Eq.~(\ref{Hsm}). We note that from Eq.~(\ref{Eq:RGg1g2})
we find that $\bar g = g_1 + g_2$ flows according to
\be
\frac{d \bar g}{d\ell} = \ve \bar g -\frac{\bar g}4(17 g_1 + 18 g_2),
\ee
ensuring that the smectic line
$\bar g = 0$ (defined by full nonlinear rotational invariance, $g_1 = - g_2$) is preserved. Examining Eqs.~(\ref{Eq:RGg1g2}), (\ref{Eq:RGdimensionless}) we further note that this Sm fixed point is stable
inside the $-g_1=g_2\equiv g$, $r=0$ subspace (the flow of $r$ reduces
to a homogeneous equation), with the flow reducing to that of a single
coupling $g$
\be
\frac{d g }{d\ell} = \ve g -\frac{5}4 g^2, \label{Eq:gflow}
\ee
and harmonic couplings
\bse
\label{Eq:BKflows}
\begin{eqnarray}
\frac{d B(\ell)}{d\ell}&=&\left(D - 1 - z + 2\chi - g(\ell)\right)B(\ell)
\;,\label{Bflow}\\
\frac{d K(\ell)}{d\ell}&=&(D-5+z + 2\chi +\oh g(\ell))K(\ell)
\;.\label{Kflow}
\end{eqnarray}
\ese
The Sm fixed point, previously studied in Refs.~\onlinecite{GolubovicWang94,LRfflo}, is
given by $g^* = 4\ve/5$, $\chi = 3\ve /5$, $\eta_B = 4\ve/5$, $\eta_K =
2\ve/5$, and $z = 2 - 3\ve/5$. As its name implies, it actually describes
strongly-fluctuating finite $T$ properties of a $D$-dimensional
smectic-A, which is thus an example of a ``critical
phase''\cite{LRfflo}.

This Sm fixed point is an $\ve$-expansion approximation for a
two-dimensional smectic. As was shown by Golubovic and
Wang\cite{GolubovicWang92,GolubovicWang94}, remarkably, the universal
exponents of a $D=2$ smectic can be obtained {\em exactly} through its
mapping onto nonequilibrium dynamics of a 1+1 dimensional
Kardar-Parisi-Zhang (KPZ) equation \cite{KardarParisiZhang86}. The
exponents for the latter were deduced to be $\chi = 1/2$, $z = 3/2$,
{\it exactly} \cite{HuseHenleyFisher85}. Curiously, an uncontrolled
one-loop approximation done directly in $D=2$ smectic also gives these
exponents exactly \cite{GolubovicWang94}.

We utilize RG flows (\ref{Eq:gflow}) and (\ref{Eq:BKflows}) to compute the long-scale smectic
phonon correlation function, finding
\begin{eqnarray}
C(\kv_\perp,k_\tau)
&\approx&\frac{T}{B(\kv) k_\tau^2 + K(\kv) k_\perp^4},
\label{Cuu}
\end{eqnarray}
with moduli $B(\kv)$ and $K(\kv)$ that are singularly
wavevector-dependent. These moduli are determined by the solutions $B(\ell)$ and $K(\ell)$ of the RG flow equations (\ref{Bflow}) and (\ref{Kflow}), with the initial conditions set by the microscopic values of $B$ and $K$.

In $D=2$ (implying $\ve=1$), the scale at which the nonlinearities become important is given by  $\xi^{NL}_\perp =
\frac{1}{T}\left(\frac{K^3}{B}\right)^{1/2}$ \cite{LRfflo}. At scales longer than $\xi^{NL}_\perp$, the nonlinear coupling
$g(\ell)$ flows to the Sm fixed point $g^*=4/5$, and the RG matching
analysis predicts anisotropic wavevector-dependent moduli
\bse
\label{KgBg}
\begin{eqnarray}
K({\bf k})&=&K\left(k_\perp\xi^{NL}_\perp\right)^{-\eta_K}
f_K\left(k_\tau\xi^{NL}_{\tau}/(k_\perp\xi^{NL}_\perp)^z\right)\;,\label{Kg}
\ \ \ \ \ \ \ \ \ \\
&\sim& k_\perp^{-\eta_K},\nonumber\\
B({\bf k})&=&B\left(k_\perp\xi^{NL}_\perp\right)^{\eta_B}
f_B\left(k_\tau\xi^{NL}_{\tau}/(k_\perp\xi^{NL}_\perp)^z\right)\;,\label{Bg}\\
&\sim& k_\perp^{\eta_B},\nonumber
\end{eqnarray}
\ese
with universal scaling functions, $f_B(x)$, $f_K(x)$ that we will not
compute here. The anomalous exponents in $D=2$ ($\ve=1$) are given by
\bse
\begin{eqnarray}
\eta_B&=&g^*={4\over 5}\;,\label{etaB2}\\
\eta_K&=&\oh g^*={2\over 5}\;,\label{etaK2}\\
z&=& 2-\oh(\eta_B+\eta_K) = \frac{7}{5}.
\end{eqnarray}
\ese
The underlying rotational invariance of the Sm fixed point gives an
{\em exact} relation between the two anomalous $\eta_{B,K}$ exponents
\bse
\begin{eqnarray}
3-D &=& {\eta_B\over 2} + {3\over  2}\eta_K\;,
\label{WI}\\
1  &=& {\eta_B\over 2} + {3\over 2}\eta_K\;,\ \ \mbox{for
  $D=2$},
\label{WI2d}
\end{eqnarray}
\ese
which is obviously satisfied by the anomalous exponents,
Eqs.~(\ref{etaB2})-(\ref{etaK2}), computed here to first order in
$\ve=3-D$. In $D=3$, this analysis reduces to the exact logarithmically renormalized $B(\bk)$ and $K(\bk)$ found by Grinstein and Pelcovits \cite{GrinsteinPelcovits82}.

\section{Summary and Conclusions \label{Sec:conclusion}}

To summarize, we studied the quantum Ising ferromagnetic transition in a one-dimensional system of itinerant electrons. Starting with a microscopic model of a quantum wire with Rashba spin-orbit coupling, we derived a bosonized effective low-energy theory that governs the transition. To analyze the theory, we used a renormalization group approach, controlled by an $\ve$-expansion. We showed that in the general case, when  inversion symmetry is absent, strong spin fluctuations necessarily drive the transition {\it first order}, in contrast to the inversion-symmetric case and the predictions of Hertz-Millis theory.

While in the present paper we consider a 1d bosonized model, we conjecture that the first-order transition is a qualitative property that extends to two- and three-dimensional itinerant ferromagnets without inversion symmetry.
This conjecture serves as a motivation for future study of the nature of quantum ferromagnetic transition in higher dimensions.

As a byproduct of our analysis, we demonstrated that the imaginary time $D=1+1$ action of the ferromagnetic wire can be mapped onto the problem of a two-dimensional smectic-A to smectic-C transition in a magnetic field. The range of parameters in the latter problem, however, is inaccessible for the problem of a ferromagnetic transition, and thus leads to qualitatively different physics. In particular, we showed that the Sm-A to Sm-C transition in two dimensions is {\it second order}, controlled by a newly found stable critical point.

Finally, we constructed the global phase diagram for a bosonic field theory that describes both ferromagnetic and Sm-A to Sm-C phase transitions. We demonstrated that a region of the first-order transition, $g_1>0$, is separated from the continuous transition, $0<-g_1<g_2$, by a tricritical point at $g_1=0$, which describes the FM transition in the presence of inversion symmetry.

{\it Acknowledgements:} LR thanks John Toner for discussions and for letting us know that he independently derived the RG equations for the SmA-to-SmC transition in a magnetic field in unpublished work. This  project  was funded by the DOE Office of Basic Energy Sciences, Division of Materials Sciences and Engineering under award DE-SC0010526 (VK and LF). JR acknowledges a fellowship from the Gordon and Betty Moore Foundation under the EPiQS initiative (grant no. GBMF4303). LR was supported by the Simons Investigator award from the Simons Foundation, by the NSF under grant no. DMR-1001240C, and by the KITP under grant no. NSF PHY-1125915.  LR thanks the KITP for its hospitality as part of the Synthetic Matter workshop and sabbatical program, when part of this work was completed.

\appendix

\section{Derivation of RG equations \label{app:diagrams}}

In this Appendix we demonstrate the derivation of RG equations~(\ref{Eq:RGset}). We focus on the FM transition which is described by the effective action (\ref{Eq:action}). To obtain the description of Sm phases, it is sufficient to substitute $B_3 \to i B_3$.

To calculate the one-loop corrections to the RG equations, we start with a bare Green's function $G_0(\omega,\bk)$,

\be
G_0(\omega, \bk) = \frac 1{B\omega^2 + r k^2 + K k^4},\label{Eq:GreenFunction}
\ee
and treat the nonlinear terms, $B_3$ and $B_4$, as a small perturbation. Next, integrating out high momenta modes in the shell $\Lambda e^{-\delta \ell} < k_> < \Lambda,$ $-\infty < \omega < \infty$, we obtain corrections to the parameters of the effective action~(\ref{Eq:action}), see Sec.~\ref{Sec:FMtransition} for details. The one-loop calculation is somewhat tedious, but straightforward.

\begin{widetext}

The correction to $B$ is given by a single diagram:

\begin{align}
 \delta B =  \raisebox{-5.5mm}{\includegraphics[width=35mm]{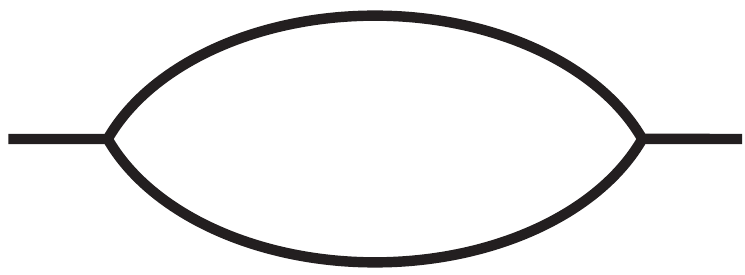}}
  &= \frac{B_3^2}2 \int_{-\infty}^{\infty}\frac{d\omega}{2\pi} \int_{\Lambda e^{-\delta \ell}}^{\Lambda} \frac{d^d k}{(2\pi)^d} \frac{k^4}{(B \omega^2 + r k^2 + K k^4)^2} \nonumber \\ &=  \frac{B_3^2}2 \int_> \frac{d\omega d^d k}{(2\pi)^{d+1}} \frac{k^4}{(B \omega^2 + r k^2 + K k^4)^2}=
 \frac{B_3^2}{B^{1/2} (K+\tr)^{3/2}} \gamma_d \delta \ell, \label{Eq:deltaB2}
\end{align}
where we defined $\tr \equiv r/\Lambda^2,$ $\gamma_d \equiv S_d \Lambda^{d-2}/8(2\pi)^d,$ and $S_d$ is the area of the sphere of unit radius in $d$ dimensions. For integer dimensions, it is given by $S_1 = 2$, $S_2 = 2\pi$, $S_3 = 4\pi$ etc. We also use a short notation $\int_>d\omega d^dk \ldots \equiv \int_{-\infty}^{\infty} d\omega \int_{\Lambda e^{-\delta\ell}}^{\Lambda} d^dk \ldots$ for a momentum shell integration hereafter.

The correction to $B_3$ is given by two diagrams:

\begin{align}
 \delta B_3^{(1)} &= \raisebox{-13mm}{\includegraphics[width=17mm]{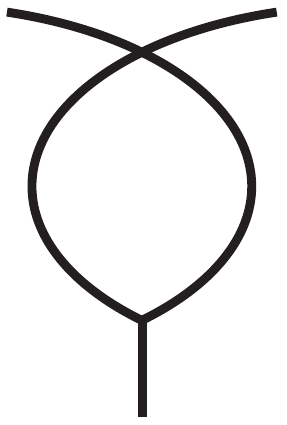}}
  = -\frac{B_3 B_4}2 \left(1 + \frac2d\right) \int_> \frac{d\omega d^d k}{(2\pi)^{d+1}} \frac{k^4}{(B \omega^2 + r k^2 + K k^4)^2}  =- \left( 1 + \frac2d \right)
 \frac{B_3 B_4}{B^{1/2} (K+\tr)^{3/2}} \gamma_d \delta \ell, \\
 \delta B_3^{(2)} &=   \raisebox{13mm}{\includegraphics[width=25mm,angle =270]{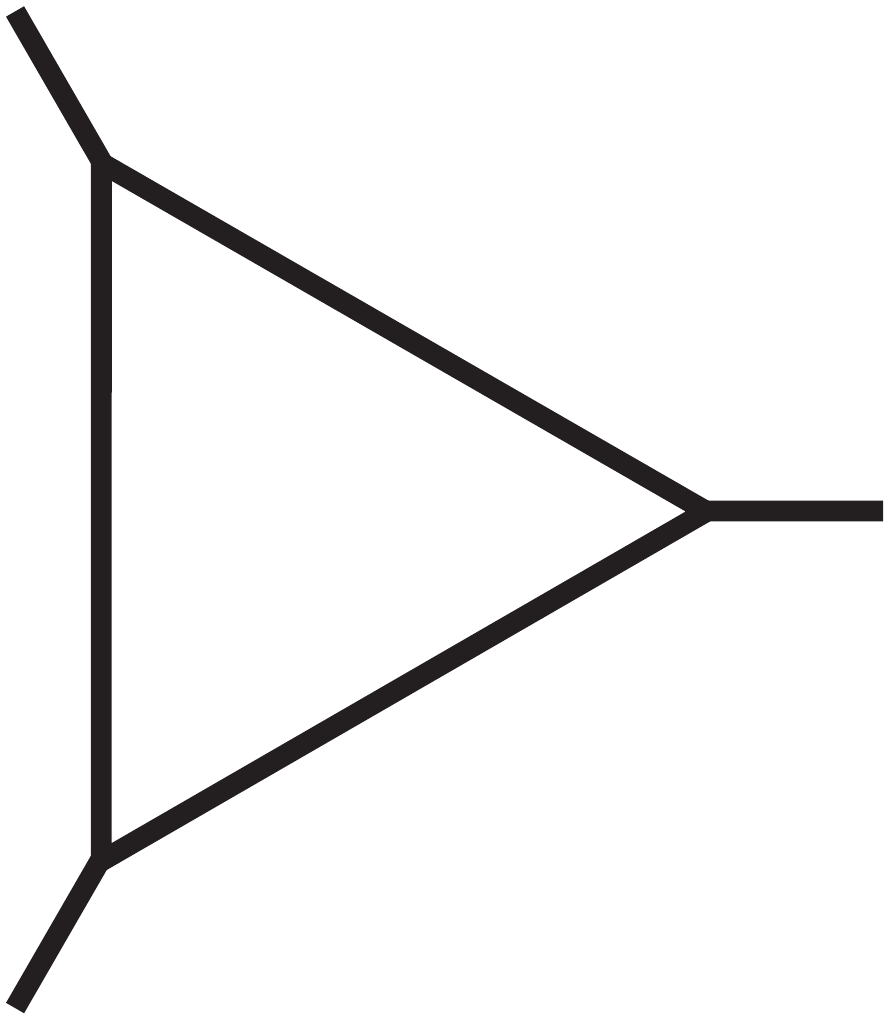}}
  = -\frac{4}d B_3^3  \int_> \frac{d\omega d^d k}{(2\pi)^{d+1}} \frac{\omega^2 k^4}{(B \omega^2 + r k^2 + K k^4)^3}  =-  \frac2d
 \frac{B_3^3}{B^{3/2} (K+\tr)^{3/2}} \gamma_d \delta \ell.
\end{align}
Summing them up, we find

\be
\delta B_3 = \delta B_3^{(1)} + \delta B_3^{(2)} = -\frac{B_3 \gamma_d \delta \ell}{B^{3/2}(K+\tr)^{3/2} d}\left[ (d+2) B_2 B_4 + 2 B_3^2 \right]. \label{Eq:deltaB3}
\ee

The correction to $B_4$ is given by three diagrams:

\begin{align}
 \delta B_4^{(1)} &= \raisebox{-9mm}{\includegraphics[width=17mm]{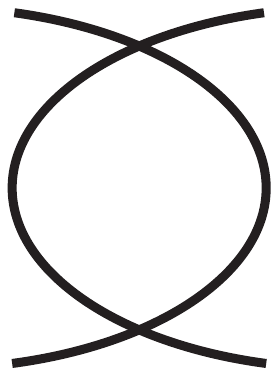}}
  = -\frac{d^2+6d + 20}{2d(d+2)}B_4^2 \int_> \frac{d\omega d^d k}{(2\pi)^{d+1}} \frac{k^4}{(B \omega^2 + r k^2 + K k^4)^2}  =- \frac{d^2 + 6d + 20}{d(d+2)}
 \frac{ B_4^2}{B^{1/2} (K+\tr)^{3/2}} \gamma_d \delta \ell, \\
 \delta B_4^{(2)} &= \raisebox{-9mm}{\includegraphics[width=30mm]{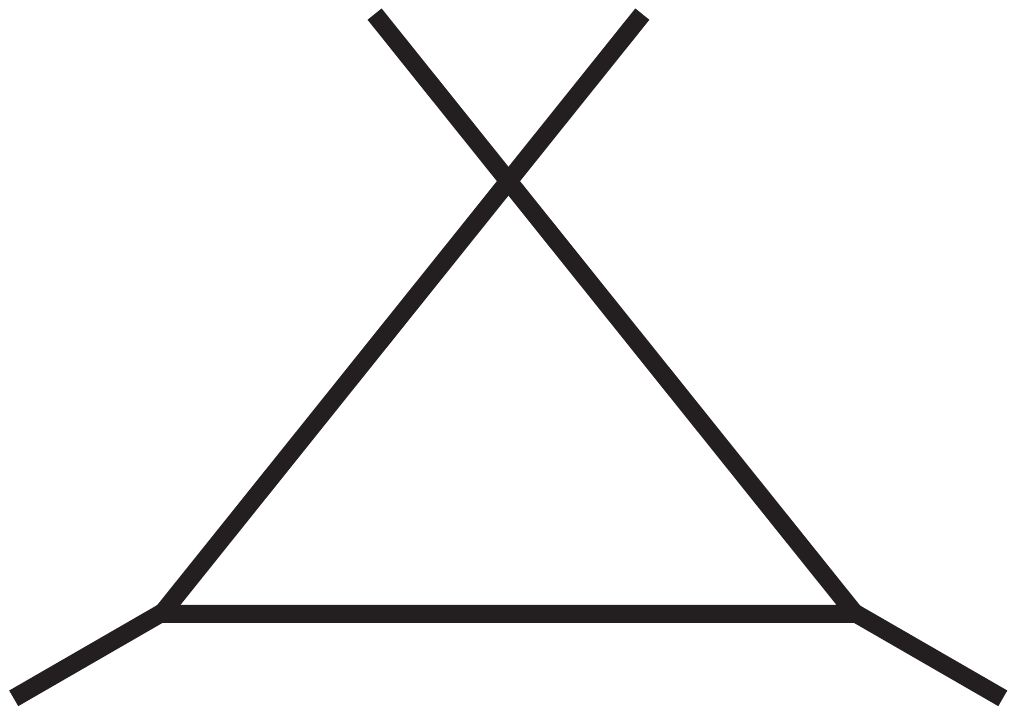}}
  = -8\frac{d+8}{d(d+2)}B_3^2 B_4 \int_> \frac{d\omega d^d k}{(2\pi)^{d+1}} \frac{\omega^2 k^4}{(B \omega^2 + r k^2 + K k^4)^3}  =- 4\frac{d+8}{d(d+2)}
 \frac{ B_3^2 B_4}{B^{3/2} (K+\tr)^{3/2}} \gamma_d \delta \ell,\\
 \delta B_4^{(3)} &= \raisebox{-12mm}{\includegraphics[width=27mm]{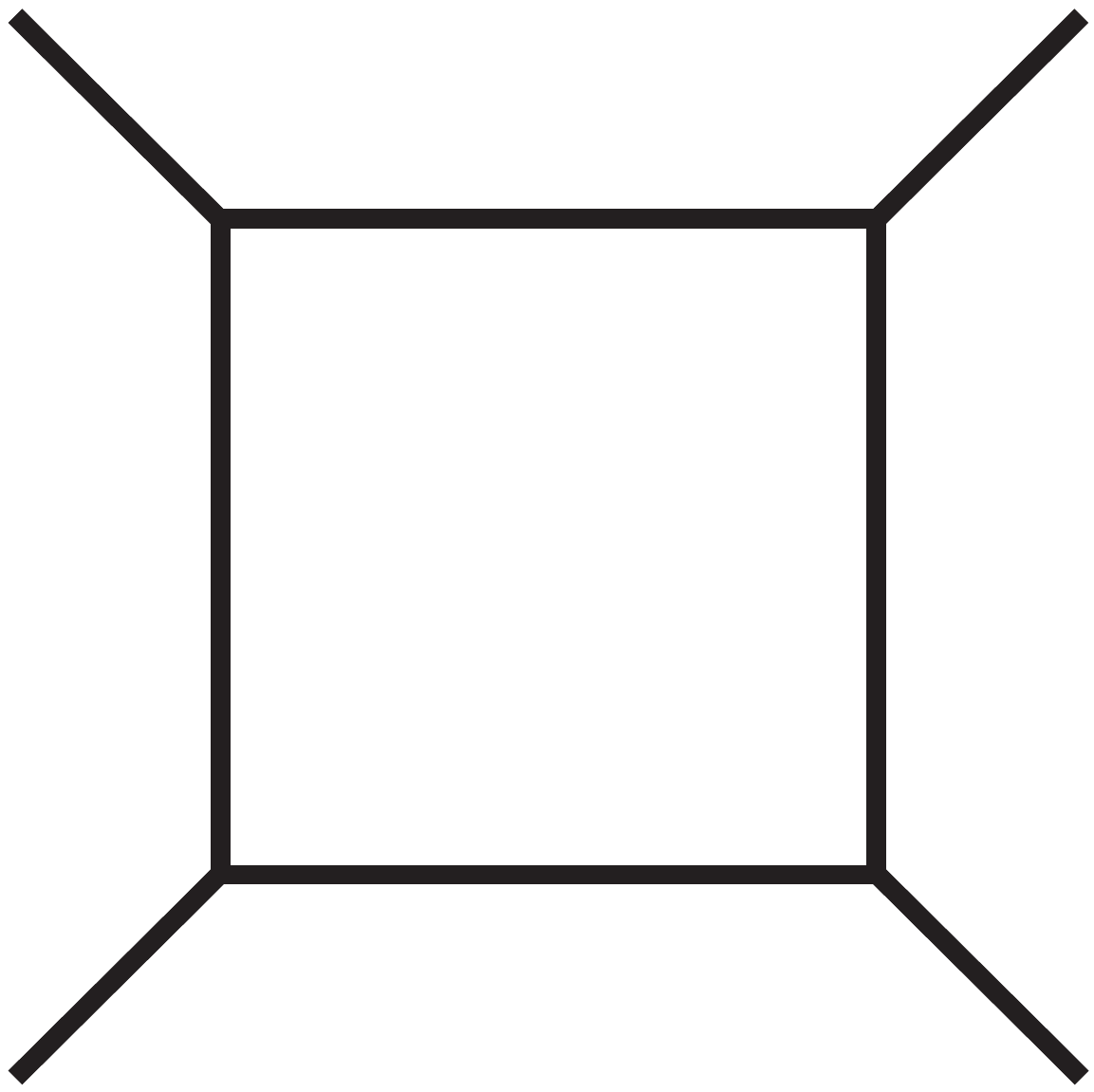}}
  = -\frac{48}{d(d+2)}B_3^4 \int_> \frac{d\omega d^d k}{(2\pi)^{d+1}} \frac{\omega^4 k^4}{(B \omega^2 + r k^2 + K k^4)^4}  =- \frac{12}{d(d+2)}
 \frac{ B_3^4}{B^{5/2} (K+\tr)^{3/2}} \gamma_d \delta \ell.
\end{align}

After summation, we find

\be
\delta B_4 = \delta B_4^{(1)} + \delta B_4^{(2)} + \delta B_4^{(3)} = -\frac{\gamma_d \delta \ell}{B^{5/2}(K+\tr)^{3/2} d (d+2)} \left[ (d^2 + 6d + 20)B^2 B_4^2 + 4(d+8) B B_4 B_3^2 + 12 B_3^4    \right]. \label{Eq:deltaB4}
\ee

The correction to $K$ is obtained from the same diagram as the correction to $B$. However, now the external legs of the diagram correspond to spatial derivatives of the field $\theta$, $\partial_x^2 \theta$, rather then time derivatives, as in case of $B$.  Furthermore, since $K$ couples to the square of the {\it second} derivative, one needs to expand the exact expression for this diagram to the next-to-leading order in slow external momentum. The result reads as

\be
\delta K = \raisebox{-5.5mm}{\includegraphics[width=35mm]{two-legcropped.pdf}} = - \frac{B_3^2 \gamma_d \delta \ell}{2 B^{3/2}(K+\tr)^{5/2}} \left[ \frac{(2K + \tr)(K + \tr)}{d} - \frac{3\tr^2}{d(d+2)} \right]. \label{Eq:deltaK}
\ee

Finally, there are three contributions to the correction to $\tr$. Two of them can be calculated directly from the one-loop diagrams:

\begin{align}
\delta \tr^{(1)} &= \raisebox{-6mm}{\includegraphics[width=22mm]{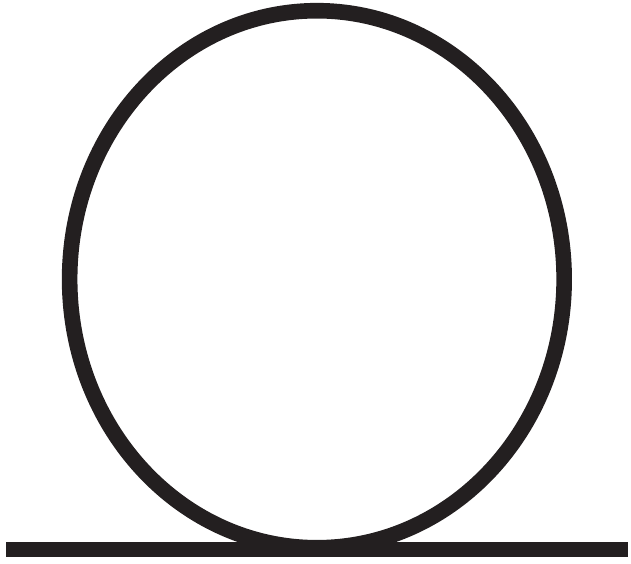}} = \frac12\left( 1+ \frac2d \right)B_4 \Lambda^{-2}\int_> \frac{d\omega d^d k}{(2\pi)^{d+1}} \frac{k^2}{B \omega^2 + r k^2 + K k^4 } = 2\left( 1 + \frac2d  \right) \frac{B_4}{B^{1/2} (K+\tr)^{1/2}} \gamma_d \delta \ell, \\
\delta \tr^{(2)} &= \raisebox{-5.5mm}{\includegraphics[width=35mm]{two-legcropped.pdf}} = \frac2d B_3^2 \Lambda^{-2} \int_> \frac{d \omega d^d k}{(2\pi)^{d+1}} \frac{\omega^2 k^2}{(B \omega^2 + r k^2 + K k^4)^2} = \frac4d \frac{B_3^2}{B^{3/2}(K+\tr)^{1/2}} \gamma_d \delta \ell.
\end{align}

To obtain third contribution, we consider the diagram that generates a new term in the effective action, $i(\delta r_\tau/2) \partial_\tau \theta$:

\be
\delta r_\tau = \raisebox{-10mm}{\includegraphics[width=18mm]{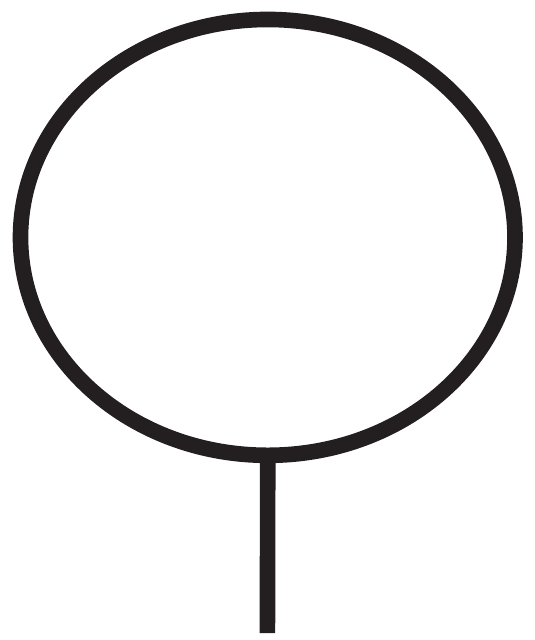}} = -B_3 \int_> \frac{d\omega d^dk}{(2\pi)^{d+1}}\frac{k^2}{B \omega^2 + r k^2 + K k^4} = -4\frac{B_3\Lambda^2}{B^{1/2} (K+\tr)^{1/2}} \gamma_d \delta \ell.
\ee
This term describes the correction to the average spin current and can be absorbed by shifting $\partial_\tau \theta \to \partial_\tau \theta - i\delta r_\tau /2B$ (or, equivalently, $\theta \to \theta - i\tau \delta r_{\tau}/2B$), such that $\partial_\tau \theta$ always describes {\it deviation} from the average spin current, i.e., $\langle \partial_\tau \theta\rangle =0.$ This extra step of RG, however, generates an additional correction to $\tr$, which reads as

\be
\delta \tr^{(3)} = -\frac{B_3 \delta r_\tau \Lambda^{-2}}{2 B} = 2\frac{B_3^2}{B^{3/2} (K+\tr)^{1/2}} \gamma_d \delta \ell.
\ee

After summation, we find

\be
\delta \tr = \delta \tr^{(1)} + \delta \tr^{(2)} + \delta \tr^{(3)} = 2\left( 1+ \frac2d \right)\frac{B B_4 + B_3^2}{B^{3/2}(K+\tr)^{1/2}} \gamma_d \delta \ell. \label{Eq:deltar}
\ee

Collecting together Eqs.~(\ref{Eq:deltaB2}), (\ref{Eq:deltaB3}), (\ref{Eq:deltaB4}), (\ref{Eq:deltaK}), and (\ref{Eq:deltar})  we exactly obtain the one-loop part of the RG set of equations (\ref{Eq:RGset}).

\end{widetext}


\begin{thebibliography}{100}
\bibitem{noteChargeSector} In fact, there are terms which couple charge and spin sectors that are relevant near the Gaussian fixed point. However, after integrating out the charge sector, these terms only contribute operators that are already present in Eq.~(\ref{Eq:action}), and thus do not change the analysis for a PM-FM transition\cite{Yang04}. Despite of this, a
  preliminary analysis shows that critical fluctuations in the spin
  sector feedback to renormalize the charge sector. In particular, we
  expect that this leads to a nontrivial $z > 1$ (rather than $z=1$ at the Gaussian fixed point) dynamical exponent for the charge sector.
\bibitem{commentKunYang} Here we corrected numerical values of critical exponents at IS critical point obtained in Ref.~\onlinecite{Yang04}.
\bibitem{commentHLM} Using duality to map a superconductor onto an XY-model it was shown\cite{DasguptaHalperin} that, in fact, (fluctuation-driven first-order transition based on $\ve=4-d$-expansion notwithstanding) a 3d normal-to-superconductor phase transition can be continuous.  This suggests a change in behavior and appearance of a tricritical point for $\ve \rightarrow 1$.
\end{thebibliography}
\end{document}